\documentstyle[12pt,aaspp4,eqsecnum,flushrt]{article}
\tighten
\lefthead{ROBERGE \& LAZARIAN}
\righthead{Davis-Greenstein Alignment}
\newcommand{\be}{\begin{equation}}
\newcommand{\ee}{\end{equation}}
\newcommand{\sameauthor}{\rule[+.5ex]{.7cm}{1pt}. \ }
\newcommand{\veca}{\mbox{\boldmath$a$}}      
      
\newcommand{\vecB}{\mbox{\boldmath$B$}}      
\newcommand{\vecE}{\mbox{\boldmath$E$}}      
\newcommand{\vecJ}{\mbox{\boldmath$J$}}

\newcommand{\ofxh}{\mbox{$\hat{\vecx}^{\bf\rm o}$}}
\newcommand{\ofyh}{\mbox{$\hat{\vecy}^{\bf\rm o}$}}
\newcommand{\ofzh}{\mbox{$\hat{\vecz}^{\bf\rm o}$}}

\newcommand{\gfxh}{\mbox{$\hat{\vecx}$}}
\newcommand{\gfyh}{\mbox{$\hat{\vecy}$}}
\newcommand{\gfzh}{\mbox{$\hat{\vecz}$}}
\newcommand{\gfbasis}{\mbox{$\gfxh,\gfyh,\gfzh$}}
\newcommand{\Jfxh}{\mbox{$\hat{\vecx}^J$}}
\newcommand{\Jfyh}{\mbox{$\hat{\vecy}^J$}}
\newcommand{\Jfzh}{\mbox{$\hat{\vecz}^J$}}
\newcommand{\Jfbasis}{\mbox{$\Jfxh,\Jfyh,\Jfzh$}}
\newcommand{\bfxh}{\mbox{$\hat{\vecx}^{b}$}}
\newcommand{\bfyh}{\mbox{$\hat{\vecy}^{b}$}}
\newcommand{\bfzh}{\mbox{$\hat{\vecz}^{b}$}}
\newcommand{\bfbasis}{\mbox{$\bfxh,\bfyh,\bfzh$}}
\newcommand{\Jcart}{\mbox{$\left(J_x,J_y,J_z\right)$}}
\newcommand{\dJcart}{\mbox{$dJ_x\,dJ_y\,dJ_z$}}
\newcommand{\vecx}{\mbox{\boldmath$x$}}      
\newcommand{\vecy}{\mbox{\boldmath$y$}}      
\newcommand{\vecz}{\mbox{\boldmath$z$}}      
\newcommand{\fv}{\mbox{$f\left(\theta,J_x,J_y,J_z\right)$}}
\newcommand{\Ts}{\mbox{$T_{\rm s}$}}
\newcommand{\Tg}{\mbox{$T_{\rm g}$}}
\newcommand{\TsTg}{\mbox{$T_{\rm s}/T_{\rm g}$}}
\newcommand{\trot}{\mbox{$t_{\rm rot}$}}
\newcommand{\tLar}{\mbox{$t_{\rm Lar}$}}
\newcommand{\tBar}{\mbox{$t_{\rm Bar}$}}
\newcommand{\tgas}{\mbox{$t_{\rm gas}$}}
\newcommand{\tmag}{\mbox{$t_{\rm mag}$}}
\newcommand{\Gpar}{\mbox{$\Gamma_{\|}$}}
\newcommand{\Gper}{\mbox{$\Gamma_{\bot}$}}
\newcommand{\Grat}{\mbox{$\Gper/\Gpar$}}
\newcommand{\Ipar}{\mbox{$I_{\|}$}}
\newcommand{\Iper}{\mbox{$I_{\bot}$}}

\newcommand{\vth}{\mbox{$v_{\rm th}$}}

\newcommand{\Jth}{\mbox{$J_{\rm th}$}}
\newcommand{\deltm}{\mbox{$\delta_{\rm m}$}}
\newcommand{\deltmI}{\mbox{$\delta_{\rm m}^{-1}$}}

\newcommand{\eps}{\mbox{$\epsilon$}}
\newcommand{\Ordeps}{\mbox{${\cal O}(\epsilon)$}}

\newcommand{\OrdmI}{\mbox{${\cal O}\left(\delta_{\rm m}^{-1}\right)$}}
\newcommand{\chip}{\mbox{$\chi^{\prime}$}}
\newcommand{\chipp}{\mbox{$\chi^{\prime\prime}$}}
\newcommand{\rhos}{\mbox{$\rho_{{\rm s}}$}}
\newcommand{\rhosz}{\mbox{$\rho_{{\rm s},0}$}}
\newcommand{\Jx}{\mbox{$J_x$}}
\newcommand{\Jy}{\mbox{$J_y$}}
\newcommand{\Jz}{\mbox{$J_z$}}
\newcommand{\Jper}{\mbox{$J_{\bot}$}}
\newcommand{\Aper}{\mbox{$A_{\bot}$}}
\newcommand{\fper}{\mbox{$f_{\bot}$}}
\newcommand{\Athet}{\mbox{$A_{\theta}$}}
\newcommand{\Bthet}{\mbox{$B_{\theta\theta}$}}

\newcommand{\QJ}{\mbox{$Q_J$}}
\newcommand{\tilQJ}{\mbox{$\tilde{Q}_J$}}
\newcommand{\dQJ}{\mbox{$\delta Q_J$}}
\newcommand{\dQJavg}{\mbox{$\overline{\delta Q_J}$}}
\newcommand{\dQJms}{\mbox{$\overline{\delta Q_J^2}$}}
\newcommand{\dQJrms}{\mbox{$\left(\overline{\delta Q_J^2}\right)^{1/2}$}}
\newcommand{\qX}{\mbox{$q_X$}}
\newcommand{\QX}{\mbox{$Q_X$}}

\newcommand{\dQX}{\mbox{$\delta Q_X$}}
\newcommand{\dQXavg}{\mbox{$\overline{\delta Q_X}$}}
\newcommand{\dQXms}{\mbox{$\overline{\delta Q_X^2}$}}
\newcommand{\dQXrms}{\mbox{$\left(\overline{\delta Q_X^2}\right)^{1/2}$}}

\newcommand{\dRms}{\mbox{$\overline{\delta R^2}$}}
\newcommand{\dRrms}{\mbox{$\left(\overline{\delta R^2}\right)^{1/2}$}}

\newcommand{\fint}{\mbox{$f_{\rm int}$}}

\newcommand{\fintv}{\mbox{$\fint\left(\theta\,|J\,\right)$}}
\newcommand{\fext}{\mbox{$f_{\rm ext}$}}
\newcommand{\vext}{\mbox{$\Jx,\Jy,\Jz$}}
\newcommand{\fextv}{\mbox{$\fext\left(\vext\right)$}}
\newcommand{\Abar}{\mbox{$\bar{A}$}}
\newcommand{\Abarx}{\mbox{$\bar{A}_x$}}
\newcommand{\Abary}{\mbox{$\bar{A}_y$}}
\newcommand{\Abarz}{\mbox{$\bar{A}_z$}}
\newcommand{\Bbar}{\mbox{$\bar{B}$}}
\newcommand{\Bbarxx}{\mbox{$\bar{B}_{xx}$}}
\newcommand{\Bbaryy}{\mbox{$\bar{B}_{yy}$}}
\newcommand{\Bbarzz}{\mbox{$\bar{B}_{zz}$}}
\newcommand{\Zfac}{\mbox{$\left[\,1+\left(h-1\right)\sin^2\theta\,\right]$}}
\newcommand{\Zthet}{\mbox{$Z(\theta)$}}

\newcommand{\Cpar}{\mbox{$C_{\|}$}}
\newcommand{\Cper}{\mbox{$C_{\bot}$}}
\newcommand{\Cavg}{\mbox{$C_{\rm avg}$}}

\newcommand{\Cxo}{\mbox{$C_x^{\rm o}$}}
\newcommand{\Cyo}{\mbox{$C_y^{\rm o}$}}

\newcommand{\Lxy}{\mbox{${\cal L}_{xy}$}}
\newcommand{\Lz}{\mbox{${\cal L}_z$}}
\newcommand{\fxy}{\mbox{$f_{xy}$}}
\newcommand{\fxyv}{\mbox{$\fxy\left(J_x,J_y|J_z\right)$}}

\newcommand{\fz}{\mbox{$f_z$}}
\newcommand{\fzv}{\mbox{$f_z\left(\Jz\right)$}}

\newcommand{\Btilde}{\mbox{$\tilde{B}$}}
\newcommand{\Btildxx}{\mbox{$\tilde{B}_{xx}$}}

\newcommand{\Btildzz}{\mbox{$\tilde{B}_{zz}$}}
\newcommand{\Atildx}{\mbox{$\tilde{A}_x$}}
\newcommand{\Atildy}{\mbox{$\tilde{A}_y$}}
\newcommand{\Atildz}{\mbox{$\tilde{A}_z$}}
\newcommand{\intf}{\mbox{$\int_{-\infty}^{+\infty}$}}
\newcommand{\vdot}{\mbox{\boldmath$\cdot$}}
\newcommand{\Dt}{\mbox{$\Delta t$}}

\begin{document}
%
%
\title{Davis-Greenstein alignment of oblate spheroidal grains}

\author{W. G. Roberge\altaffilmark{1} \& A. Lazarian\altaffilmark{2}}
\altaffiltext{1}{Dept.\ of Physics, Applied Physics \& Astronomy,
Rensselaer Polytechnic Institute, Troy, NY 12180, USA, roberw@rpi.edu}
\altaffiltext{2}{Dept.\ of Astrophysical Sciences, Princeton University;
Current address: Canadian Institute for Theoretical Astrophysics,
University of Toronto, Toronto, Ontario, Canada M5S 3H8;
lazarian@cita.utoronto.ca}

\author{to appear in {\it Monthly Notices of the Royal Astronomical Society}}

\begin{abstract}
We present extensive calculations on the efficiency of grain
alignment by the Davis-Greenstein mechanism.
We model the grains as oblate spheroids with arbitrary axis
ratios.
Our description of the grain dynamics includes
(i) magnetic dissipation and the inverse
process driven by thermal fluctuations in the grain magnetization;
(ii) gas-grain collisions and thermal evaporation of molecules
from the grain surface;
(iii) the transformation of rotational energy into heat by the Barnett
effect and the inverse process driven by thermal fluctuations; and
(iv) rapid Larmor precession of the grain angular momentum about the
interstellar magnetic field.
For ordinary paramagnetic grains, we calculate the Rayleigh reduction
factor, $R$, for $>$$1000$ combinations of the 3 dimensionless parameters
which characterise the alignment.
For superparamagnetic grains, we calculate $R$ from an exact
analytic solution for the relevant distribution function.
Our results are compared with classical studies of DG alignment,
which did not include the Barnett effect.
We calibrate the accuracy of a  recently-proposed perturbative
approximation, which includes the Barnett effect, and show that
it yields $R$ values with a mean error of $\approx$$17$\%.
\end{abstract}

\setcounter{footnote}{0}

%
%
\section{Introduction}

Nearly half a century after the discovery of interstellar polarization,
the alignment of interstellar dust grains remains poorly understood.
One candidate for the mysterious alignment mechanism is a process
described by Davis and Greenstein (1951, henceforth DG51), in which
paramagnetic dissipation aligns the grain angular momentum, \vecJ,
%
%
with the interstellar magnetic field, \vecB. The alignment of the
grain {\it axes}\/ with \vecB\ requires a correlation between
\vecJ\ and the axial directions. In the original formulation of
DG51, the required correlation is produced by the torques due to
paramagnetic dissipation and gas damping.
%
The Davis-Greenstein (DG) mechanism was subsequently modified to describe
ferromagnetic grains (Spitzer \& Tukey 1951) and
to include the important disorienting effects of thermal
fluctuations in the magnetization (Jones \& Spitzer 1967, henceforth JS67).
The ``classical age'' of grain alignment research 
culminated with extensive quantitive investigations
based on the Fokker-Planck equation (JS67) and
Monte Carlo methods (Purcell 1969; Purcell \& Spitzer 1971,
henceforth PS71).
These studies concluded that the DG mechanism
can explain the polarization observed toward diffuse clouds
if the grains are superparamagnetic, but requires magnetic fields
$>10^{-5}$\,G if the grains are composed of ordinary paramagnetic
substances.

The classical era was followed by a twenty-year period with little
research on the DG effect, during which interest focussed
on an alternative mechanism proposed by Purcell (1979, henceforth P79).
In Purcell's mechanism, magnetic dissipation aligns the angular
momenta with \vecB, as in DG51, but rotation at superthermal energies
stabilizes the grains against disorientation by the gas particles.
P79 also showed that the ``internal alignment'' of the grain
axes with \vecJ\ is determined by the conversion of rotational
energy into heat via frictional processes internal to the grain
%
(e.g., the Barnett effect).
%
Internal dissipation drives an isolated grain toward the state
of minimum energy consistent with angular momentum conservation.
For rotational energies much larger than the equipartition
energy, the result is almost perfect alignment between \vecJ\ and
the principal axis of largest rotational inertia.

Research on the DG mechanism has been reinvigorated by studies
of polarization toward molecular clouds, where the
physical conditions appear unfavorable for Purcell's mechanism 
(and other mechanisms, see Hildebrand 1988, 1996 and references therein).
Roberge, DeGraff and Flaherty (1993, henceforth RDGF93) modified
the DG effect to include internal dissipation in the regime where
the internal alignment is perfect. Lazarian (1995, henceforth L95)
presented an accurate analytic solution for the alignment efficiency
in this regime.
Detailed studies of internal dissipation
(Lazarian 1994; Lazarian \& Roberge 1997, henceforth L94 and LR97,
respectively) showed that internal alignment becomes imperfect at
thermal rotation energies and produced an accurate quantitative theory
of the dissipation mechanism.
The modification of the DG effect to include imperfect internal
alignment was carried out by Lazarian (1997, henceforth L97), who
gave an approximate analytic solution of the Fokker-Planck equation.

The approach of L97 replaces certain random variables in the Fokker-Planck
equation with their mean values. Though physically well motivated,
this is an ad hoc approximation whose accuracy can only be determined
by comparison with calculations of known accuracy. In this paper we
present such calculations.  Our plan is as follows:
The dynamical processes and modelling assumptions included
in our calculations are summarised in \S2.
In \S3 we describe an adiabatic approximation
(Roberge 1997, henceforth R97) that yields accurate solutions
of the Fokker-Planck equation along with a rigorous procedure
for estimating the errors.
In \S4 we evaluate these solutions numerically for
ordinary paramagnetic grains (\S4.1) and analytically for
superparamagnetic grains (\S4.2).
We compare our results with the classical studies of DG
alignment in \S5.1 and with the results of L97 in \S5.2.
The status of the DG mechanism, including outstanding problems,
is discussed in \S6 and our results are summarised in \S7.

%
%
\section{Modelling assumptions}

\subsection{Grain model}

We model the grains as oblate spheroids with semiaxes $a$ parallel to the
symmetry axis and $b$ perpendicular to the symmetry axis.
Let $\left\{\bfbasis\right\}$ be the Cartesian basis of a ``body frame''
attached to
the grain, where \bfzh\ is parallel to the symmetry axis, \veca\ (Fig.~1).
The components of the inertia tensor for rotation parallel and
perpendicular to \bfzh\ will be denoted \Ipar\ and \Iper, respectively.
Our model grains are composed of an unspecified solid with uniform
density \rhos, temperature \Ts, and frequency-dependent magnetic
susceptibility
$\chi\left(\omega\right)=\chip\left(\omega\right)+
\imath\chipp\left(\omega\right)$.
They are located in a static gas of particles with mass $m$,
number density $n$, and kinetic temperature \Tg, plus a uniform magnetic
field, \vecB.

\subsection{Dynamical processes}

We are interested ultimately in the distribution of the grain
orientations in an inertial frame fixed to the gas.
Let $\left\{\gfbasis\right\}$  be the basis of the inertial frame,
with \gfzh\ parallel to \vecB\ and the other basis vectors
oriented arbitrarily in the plane perpendicular to \vecB\ (Fig.~2).
The orientation of \veca\ in the inertial frame changes due to
(i) the motion of the grain's axes with respect to
\vecJ\ caused by rotation and the Barnett effect,
plus (ii) the motion of \vecJ\ in the inertial frame caused by
%
Larmor precession and the DG effect.
%
It is therefore natural to specify the orientation in 
terms of the four angles defined in Figures~1 and 2, so that
$\left(\theta,\psi\right)$ gives the orientation of \veca\ with respect to
\vecJ\ and $\left(\beta,\phi\right)$ gives the orientation of
\vecJ\ with respect to \vecB.

The dynamical timescales for these variables are listed in Table~1.
The rotational motion of a spheroid causes \veca\ to precess about
\vecJ\ with period $\sim\trot$ (for reasonable grain shapes)
and the interaction of the grain magnetic moment with the interstellar
magnetic field causes \vecJ\ to precess about \vecB\ with  period \tLar.
The smallness of the rotation and Larmor periods compared to
all other timescales of interest insures that
$\psi$ and $\phi$ are uniformly distributed.
Thus, only the joint distribution of $\theta$ and $\beta$ needs
to be calculated in practice.

We assume that internal dissipation is dominated by the Barnett
effect\footnote{A recent study by Lazarian \& Efroimsky (1999)
shows that, for grains with large axis ratios and grains produced
by agglomeration, internal dissipation is dominated by inelasticity.
However, this has no effect on the conclusions of the present study,
which only requires the timescale for internal dissipation to be
much smaller than the timescale for external interactions.},
which therefore determines the the motion of
$\theta$.\footnote{In principle, we should also include gas damping:
the impulsive change in \vecJ\ due to a gas-grain collision
or evaporation from the grain surface changes $\theta$.
In omitting these effects we introduce errors in the $\theta$ distribution
that, for typical molecular clouds, are of order $\tBar/\tgas \sim 10^{-3}$.}
The systematic and random motions of $\theta$ produced by
Barnett dissipation and thermal fluctuations in the Barnett magnetization
(collectively termed ``Barnett relaxation'') have dynamical timescale
\be
\tBar = {\mu^2I_{\parallel}^3 \over VKh^2(h-1)J^2}
\label{eq-2_1}
\ee
%
(RDGF93),
%
where $V$ is the grain volume,
$h\equiv\Ipar/\Iper$, $K\equiv \chipp/\omega$ 
and $\mu$ is the magnetogyric
ratio of the orientable spins or orbits responsible for the Barnett
magnetization.
The Barnett time depends on the angular momentum; the
value given in Table~1 was computed for a grain with $J$ equal to
\be
\Jth \equiv \sqrt{\Ipar k\Tg},
\label{eq-2_2}
\ee
a typical value for thermal rotation.

We assume that the motion of $\beta$ is caused by 
(i) the systematic and random torques produced by gas-grain
collisions and thermal evaporation of molecules from the grain
surface (``gas damping''); and
(ii) the systematic and random torques produced by paramagnetic
or superparamagnetic relaxation (``magnetic damping'').
The gas damping time is the timescale for \vecJ\ to be
reduced by rotational friction with the gas.
For a spheroid rotating about its symmetry axis,
\be
\tgas \equiv
\frac{3I_{\parallel}}{4\sqrt{\pi}nmb^{4}v_{\rm th}\Gpar}
\label{eq-2_3}
\ee
(RDGF93),
where $\vth\equiv\sqrt{2kT_{\rm g}/m}$ is the gas thermal speed
and $\Gpar$ is a dimensionless coefficient that depends weakly on
the grain shape [see eq.\ (\ref{eq-A_7})].
The magnetic damping time for a spheroid rotating about its symmetry axis
is
\be
\tmag \equiv {I_{\parallel} \over KVB^2}.
\label{eq-2_4}
\ee
The choice of a particular rotation axis for the purpose of
defining characteristic timescales obviously involves no loss of generality,
but must be taken into account when comparing results from different
papers.
In particular, PS71 defined \tgas\ and \tmag\ with respect to rotation
about the {\it transverse}\/ axis. 
To compare our results with those of PS71, it is only necessary to
note that $\deltm=\left(\Gper/\Gpar\right)\,\delta$, where
\Gper\ is a dimensionless function of the grain shape
[see eq.~(\ref{eq-A_8})], \deltm\
is our magnetic damping parameter [see eq.\ (\ref{eq-3_17})]
and $\delta$ is the magnetic damping parameter of PS71.

\subsection{Characterisation of the alignment}

Our objective is to calculate the quantity which characterises
the mean axial alignment for an ensemble of identical grains.
Because different definitions of this quantity have appeared in the
literature, due to differing assumptions about the
grain dynamics, a brief review of this subject is appropriate.

Consider an electromagnetic wave propagating along the \ofzh\ axis
of the Cartesian ``observer frame'' defined in Figure~3.
The transfer equations for the Stokes parameters
depend on the cross sections\footnote{We use
the term ``cross section'' generically to refer to the extinction and
phase advance cross sections.} \Cxo\ and \Cyo\ for linearly polarized
waves with the electric vector, \vecE, along the \ofxh\ and \ofyh\ directions
(e.g., see Martin 1974, Lee \& Draine 1985).
To calculate these ``observer frame'' cross sections,
one transforms the components of \vecE\ to
a frame aligned with the principal axes of the grain and
takes the appropriately-weighted sum of the
cross sections, \Cpar\ and \Cper, for \vecE\ polarized along the grain
axes.
When the transformation is carried out and the resulting
expressions are averaged over $\psi$ and $\phi$, one finds that
the mean cross sections for an ensemble of oblate spheroids are
\be
\Cxo = \Cavg + \frac{1}{3}\,R\,\left(\Cper-\Cpar\right)\,
       \left(1-3\cos^2\zeta\right)
\label{eq-2_5}
\ee
and
\be
\Cyo = \Cavg + \frac{1}{3}\,R\,\left(\Cper-\Cpar\right),
\label{eq-2_6}
\ee
where $\Cavg\equiv\left(2\Cper+\Cpar\right)/3$ is the effective
cross section for randomly-oriented grains.
Grain alignment is characterised by the Rayleigh reduction factor, 
\be 
R \equiv \left<G\left(\cos^2\theta\right)\,G\left(\cos^2\beta\right)\right>,
\label{eq-2_7}
\ee
where $\theta$ is the angle between the axis of largest rotational inertia
(i.e., \bfzh\ for a homogeneous oblate spheroid),
$\beta$ is the angle between \vecJ\ and \vecB,
\be
G(x) \equiv \frac{3}{2}\left(x-\frac{1}{3}\right)
\label{eq-2_8}
\ee
and $\left<\right>$ denotes the ensemble average.
Thus, our objective is to calculate $R$.

Two related ``alignment measures'' which frequently appear in the literature are
\be
Q_X \equiv \left<G\left(\cos^2\theta\right)\right> =
\frac{3}{2}\left( \left<\cos^2\theta\right>-\frac{1}{3} \right),
\label{eq-2_9}
\ee
which measures the mean alignment of \veca\ with \vecJ, and
\be
Q_J \equiv \left<G\left(\cos^2\beta\right)\right> =
\frac{3}{2}\left( \left<\cos^2\beta\right>-\frac{1}{3} \right),
\label{eq-2_10}
\ee
which describes the alignment of \vecJ\ with \vecB.
The Rayleigh reduction factor cannot be determined uniquely from
%
\QJ\ and \QX\ without making additional approximations because
in general $\theta$ and $\beta$ are correlated.
%
For example, if one assumes that Barnett dissipation aligns
\veca\ {\it perfectly}\/ with \vecJ\ 
(e.g., see Lee \& Draine 1985; RDGF93; L95),
then $G(\theta)=1$ and $R=\QJ$.
This assumption is appropriate when the grains rotate
with energies $\gg k\Ts$ (P79) but a poor approximation
for thermal rotation, the case of interest here (L94; LR97).
We make no {\it a priori}\/ assumptions
about a possible relationship between $R$, \QX, and \QJ.

%
%
\section{Mathematical formulation}

\subsection{The Fokker-Planck equation}

Although we are interested only in the distribution
of $\theta$ and $\beta$, it is easier for practical reasons to
calculate the joint distribution of $\theta$ and all 3 angular momentum
coordinates.
Let \Jcart\ be the Cartesian coordinates of \vecJ\ in the inertial
frame.
Define the distribution function, \fv, so that
$\fv\,d\theta$\,$dJ_x\,dJ_y\,dJ_z$ is the joint probability of finding
\veca\ oriented in the infinitesimal interval $d\theta$ centered at
$\theta$ and the angular momentum in \dJcart\ centered at \Jcart.
Then $f$ satisfies the Fokker-Planck equation,
\be
{\partial f \over \partial t} =
-{\partial\over\partial\theta}\,\left[\,
\Athet f - {\partial\over\partial\theta}\left(\Bthet f\right)
\,\right]
-{\partial\over\partial J_k}\,\left[\,
A_k f - {\partial\over\partial J_k}\left(B_{kk} f\right)
\,\right],
\label{eq-3_1}
\ee
where repeated indices are summed over $x,y,z$ and we have
anticipated the result [see eqs.~(\ref{eq-3_19})--(\ref{eq-3_21})] that 
$B$ is a diagonal tensor.
The diffusion coefficients \Athet\ and \Bthet, which describe the
variations in $\theta$ due to Barnett relaxation, were derived in LR97.
However they are not needed in the approximation described below (see \S3.2).
The coefficients $A_k$ and $B_{kk}$ describe
the analogous changes in the angular momentum components
due to the combined effects of gas- and magnetic damping.
They are given for reference in Appendix~A.
The angular momentum diffusion coefficients are functions of
$\theta$, $\left\{J_k\right\}$ and
physical parameters such as the grain temperature and
magnetic field strength.
We assume that all physical parameters are constant over times
much longer than the timescales in Table~1, so that
only the steady solutions of eq.~(\ref{eq-3_1}) are of interest.

\subsection{Adiabatic elimination of \boldmath$\theta$}

According to Table~1, the dynamical timescale for $\theta$
is typically $\sim$3 orders of magnitude shorter than the
timescales for the angular momentum components.\footnote{There is
no need to modify this statement if the grains are superparamagnetic,
since superparamagnetism would reduce \tmag\ and \tBar\
by the same factor.}
It is possible to
exploit this disparity to greatly simplify the calculation
of $f$.
In particular, R97 showed that
\be
\fv = \fintv\,\fextv + \Ordeps,
\label{eq-3_2}
\ee
where \fintv\ can be calculated from equilibrium
thermodynamics,\footnote{It follows that \fint\ is independent, to
\Ordeps, of the particulars of the relaxation mechanism.}
\fext\ satisfies a simplified Fokker-Planck equation described
below and
$\eps\sim 10^{-3}$ is the ratio of \tBar\ to the shortest
timescale for the motion of \vecJ.

LR97 showed that the conditional $\theta$ distribution is
\be
\fintv = C_{\theta}\,\sin\theta\,\exp\left(-\xi^2\sin^2\theta\right),
\label{eq-3_3}
\ee
where $C_{\theta}$ is a normalization constant and
\be
\xi^2 \equiv {\left(h-1\right)J^2 \over 2\Ipar k\Ts}.
\label{eq-3_4}
\ee
Expression (\ref{eq-3_3}) is
the thermal equilibrium $\theta$ distribution
for an isolated grain with fixed angular momentum.
It depends parametrically on $J$ as the notation \fintv\ suggests.
This result has an obvious physical interpretation:
as the angular momentum slowly changes, the rapid evolution
of $\theta$ allows \fint\ to relax adiabatically to the
equilibrium distribution 
that would obtain if $J$ were frozen 
at its instantaneous value.
Consequently, we refer to expression (\ref{eq-3_2}) as the
adiabatic approximation.

The statistics of \fint\ are functions of $J^2$.
Examples that appear frequently in the following discussion are
the conditional mean values of $\sin^2\theta$,
\be
\sigma \equiv \int_0^{\pi}\,\sin^2\theta\,\fintv\,d\theta,
\label{eq-3_5}
\ee
$\cos^2\theta$,
\be
\kappa \equiv 1 - \sigma,
\label{eq-3_6}
\ee
and the related alignment measure
\be
q_X \equiv \frac{3}{2}\left(\kappa-\frac{1}{3}\right).
\label{eq-3_7}
\ee
They are plotted in Figure~4.
The alignment of \veca\ with \vecJ\ is different for each grain
in the ensemble, with $q_X$ increasing from zero
for grains with $J=0$ to unity for grains with $J^2 \gg (h-1)\Ipar k\Ts$.
The mean alignment for the ensemble is described by \QX, 
the average of $q_X$ over the angular momentum distribution.

In the adiabatic approximation, the angular momentum distribution
satisfies the Fokker-Planck equation
\be
-{\partial\over\partial J_k}\,
\left[\,\bar{A}_k\,\fext\,\right]
+
{1 \over 2}\,{\partial^2\over\partial J_k^2}\,
\left[\,\bar{B}_{kk}\,\fext\,\right]=0,
\label{eq-3_8}
\ee
where the coefficients $\bar{A}_k$ and $\bar{B}_{kk}$ are
obtained by averaging the $\theta$-dependent coefficients
in expression (\ref{eq-3_1}) in the obvious way. Thus,
\be
\bar{A}_k \equiv \int_0^{\pi}\ d\theta\,A_k\,\fintv
\label{eq-3_9}
\ee
and
\be
\bar{B}_{kk} \equiv \int_0^{\pi}\ d\theta\,B_{kk}\,\fintv.
\label{eq-3_10}
\ee
With $\theta$ thus eliminated, the grain alignment problem
reduces to solving equation (\ref{eq-3_8}) for the angular
momentum distribution.
Once this is accomplished, the Rayleigh reduction factor can be found by
evaluating
\be
R = \int\ d^3J\,\fextv\,q_X\left(J^2\right)\,G\left(\cos^2\beta\right),
\label{eq-3_11}
\ee
where $\int$ denotes the definite integral over all
of angular momentum space.

\subsection{Dimensionless parameters}

We minimize the number of independent parameters in the problem
by writing \Abar\ and \Bbar\ in terms of dimensionless units, with
angular momentum measured in units of \Jth\ and
time measured in units of \tgas.
Henceforth it is implied that all quantities are in dimensionless
units unless it is explicitly stated otherwise.

After evaluating expression (\ref{eq-3_9}) and writing the result
in dimensionless form, one finds that the mean torque has components
\be
\bar{A}_x= -k_{xy}\,J_x,
\label{eq-3_12}
\ee
\be
\bar{A}_y= -k_{xy}\,J_y
\label{eq-3_13}
\ee
and
\be
\bar{A}_z= -k_{z}\,J_z.
\label{eq-3_14}
\ee
The damping rates,
\be
k_{xy} \equiv 1+\deltm+\left[\lambda+(h-1)\deltm\right]\sigma
\label{eq-3_15}
\ee
and
\be
k_{z} \equiv 1+\lambda\sigma,
\label{eq-3_16}
\ee
%
are functions of $J^2$ (through $\sigma$).
%
They depend also on the magnetic damping parameter,
\be
\deltm \equiv {t_{\rm gas} \over t_{\rm mag}},
\label{eq-3_17}
\ee
and the dimensionless shape parameters $h-1$ and
\be
\lambda \equiv h\Gper/\Gpar-1
\label{eq-3_18}
\ee
(see Fig.~5).

The evaluation of the diffusion tensor yields
\be
\bar{B}_{xx} =
\left(1+\TsTg\right)
\left(1-\gamma\eta_{xy}\right)+2\left(\TsTg\right)\deltm,
\label{eq-3_19}
\ee
\be
\bar{B}_{yy} = \bar{B}_{xx}
\label{eq-3_20}
\ee
and
\be
\bar{B}_{zz} = 
\left(1+\TsTg\right)
\left(1-\gamma\eta_z\right)+2\left(\TsTg\right)\deltm,
\label{eq-3_21}
\ee
where
\be
\gamma \equiv 1-\Gper/\Gpar
\label{eq-3_22}
\ee
is a dimensionless function of the grain shape (Fig.~5).
The diffusion tensor depends on $J^2$ and $\cos^2\beta$ via the coefficients
\be
\eta_{xy}\left(J^2,\cos^2\beta\right) \equiv \frac{1}{2}\left[1+\cos^2\beta
+\frac{1}{2}\sigma\left(1-3\cos^2\beta\right)\right]
\label{eq-3_23}
\ee
and
\be
\eta_z\left(J^2,\cos^2\beta\right)
\equiv 1-\cos^2\beta-\frac{1}{2}\sigma\left(1-3\cos^2\beta\right).
\label{eq-3_24}
\ee
We conclude that the diffusion coefficients, and therefore $R$,
depend on \TsTg, \deltm\ and 3 dimensionless functions of the
grain shape.
We will parameterize our solutions by \TsTg, \deltm\ and $a/b$.

%
%
\section{Results}

\subsection{Ordinary paramagnetic grains}

Equation (\ref{eq-3_8}) must be solved numerically in general.
We do this using an algorithm described  in RDGF93, to which we refer the
reader for a thorough description of the method.
In brief, we exploit the mathematical equivalence of
eq.\ (\ref{eq-3_8}) and the coupled Langevin equations
\be
dJ_k = \bar{A}_k\,dt\ +\surd \bar{B}_{kk}\,dW_k \ \ \ \ \mbox{(no summation on $k$)},
\label{eq-4_1}
\ee
where $\left\{dW_k\right\}$ are independent Gaussian random
variables with variance $dt$.
We integrate these equations numerically as an initial value problem
to generate simulations of the time-dependent angular momentum components
and compute statistics of the angular momentum distribution, such as $R$, 
by equating time- and ensemble averages.
Theorems insure that our statistics converge to their exact values
in the limits
$\Delta t \rightarrow 0$ and $T \rightarrow \infty$, where
$\Delta t$ is the time step and $T$ is the total averaging time.
The benchmark tests and ``production'' calculations  reported in
this paper use $\Delta t = 10^{-3}$ and $T=10^5$ dimensionless units
unless it is stated otherwise.

Because we are interested only in statistics of the steady-state
distribution, the initial conditions on eq.\ (\ref{eq-4_1})
are arbitrary.
Our results were  obtained by initially setting the magnitude of \vecJ\ to
unity and choosing its direction randomly from an isotropic distribution.
To allow our simulated grains to ``forget'' these initial conditions,
we integrated the Langevin equations for 1000 dimensionless time units
before commencing our calculation of the time averages.

Useful checks on our numerics are provided by special cases where
analytic solutions for certain statistics exist.
For example, \QJ\ can be calculated exactly for the special
case $a/b=1$ (i.e., for spherical grains; see JS67, PS71).
Let \tilQJ\ (a random variable) denote the estimate of \QJ\ predicted
by one numerical integration with $T=10^5$ and $\dQJ\equiv\tilQJ-\QJ$ be the
corresponding error.
In Figure~6 we plot the error distribution obtained from
100 numerical integrations with $a/b=1$ and the other parameters set
arbitrarily to $\deltm=1$ and $\TsTg=0.5$.
The mean and rms errors are $\dQJavg= -1.5 \times 10^{-4}$ and
$\dQJrms= 1.2\times 10^{-3}$, respectively.
We infer that the uncertainty in \QJ\ is dominated by statistical
fluctuations in the simulations rather than systematic errors in
the numerics.

In the limit $\TsTg\rightarrow 0$, the alignment of \veca\ with
\vecJ\ becomes perfect and $R\rightarrow\QJ$.
L95 developed a perturbative approximation
which calculates $\QJ$ in this regime.
Figure~7 compares results obtained with our numerical method
for the case $\TsTg=0$, $a/b=2/3$ (symbols) and results obtained
with the perturbative approximation in fifth order (solid curve).
The mean and rms discrepancies between the 49 numerical
points and the corresponding predictions of perturbation theory
are $\dQJavg= 1.2 \times 10^{-2}$ and $\dQJrms= 1.3\times 10^{-2}$.
These discrepancies should be interpreted as upper limits on the
errors in the numerical results, since they are due to the combined
effects of errors in the numerics and the finite accuracy of
the perturbation method.

Another check on the numerics is provided by the case $\TsTg=1$, where the
angular momentum distribution is
Maxwellian and \QX\ can be calculated exactly (see JS67).
Figure~8 shows the errors in \QX\ for 100 trials
with $\TsTg=1$ and the other parameters
arbitrarily chosen to be $a/b=0.5$ and $\deltm=1$.
The mean and rms errors are $\dQXavg= -5\times 10^{-4}$ and
$\dQXrms= 5\times 10^{-4}$, respectively.


There do not appear to be any benchmarks that would permit us
to calibrate the errors in $R$--- the quantity of ultimate interest---
for more general cases.
However rough estimates can be obtained by assuming
(albeit somewhat unrealistically, see the discussion below) that $\theta$ and
$\beta$ are uncorrelated.
With this assumption, one can easily show that the mean square
error in $R$ satisfies
\be
\dRms = Q_J^2 \times \dQXms\ +\ Q_X^2 \times \dQJms \ + \ {\rm terms},
\label{eq-4_2}
\ee
where the omitted terms involve higher-order products of the small
quantities \dQJ\ and \dQX.
If we use the fact that \QJ\ and \QX\ cannot exceed unity and
assume, reasonably, that the values of \dQJms\ and \dQXms\ from
the benchmarks above are typical, then
\be
\dRrms \lesssim \ 1 \times 10^{-3}.
\label{eq-4_3}
\ee
Thus, we estimate a typical error in $R$ to be a few units
in the third decimal place.

We have computed the alignment of ordinary paramagnetic grains
for 1259 parameter combinations on the uniform grids
$a/b=0.1\,(0.2)\,0.9$, $\TsTg=0\,(0.1)\,0.9$, and $\log\deltm=-1\,(0.1)\,1.5$.
Our results are given in Tables~2--6.\footnote{
These data are also available in electronic form; inquiries should be
directed to WR.}
We have deleted the table entries whenever $R < 5\times 10^{-3}$, consistent
with our error estimate.
Davis-Greenstein alignment with $R$ values in the omitted range may be
relevant to polarization in the 2175\,\AA\ extinction feature,
which has been attributed to magnetically-aligned graphite grains with
$R=1$--$2\times 10^{-3}$ (Wolff et al.\ 1997).
This hypothesis is examined in a forthcoming paper
(Roberge \& Karcz, in preparation), where DG alignment of graphite
grains is studied using perturbation theory.

The dependence of various alignment measures on the parameters is shown
in Figures~9--11.
The effects of Barnett relaxation on the grain dynamics are illustrated by
the graphs of \QX.
Barnett dissipation generally enhances the alignment
of \veca\ with \vecJ, in the
sense that \QX\ is typically several times larger
than the \QX\ value for a Maxwellian distribution.
Thermal fluctuations in the Barnett magnetization account for the
systematic dependence of \QX\ on the parameters.
These fluctuations cause random changes $\sim k\Ts$ in the rotational energy,
\be
E_{\rm rot} = {J^2\over 2I_{\parallel}}
\left[1+(h-1)\sin^2\theta\right].
\label{eq-4_4}
\ee
The resulting variations in $\theta$ inhibit the alignment of \veca\ with \vecJ\ unless
\be
{2I_{\parallel}kT_{\rm s} \over (h-1)J^2 } \ll 1,
\label{eq-4_5}
\ee
that is, unless $k\Ts$ is much smaller than the difference between
the minimum ($\theta=0$) and maximum ($\theta=\pi/2$) rotational energies.
It follows from equation (\ref{eq-4_5}) that \QX\ should decrease as the grains become
more spherical (so that $h\rightarrow 1$),
as the magnetic damping parameter increases (so that the mean value of $J^2$ decreases)
and as the dust temperature increases.
These trends are present in Figures 9, 10 and 11,
respectively.
The graphs of \QX\ also provide additional checks on our calculations.
Thus, \QX\ vanishes in the limit $a/b \rightarrow 1$ (spherical grains
disoriented completely by an infinitesimal fluctuation) and
\QX\ approaches the Maxwellian solution as $\TsTg\rightarrow 1$.

The graphs of \QJ\ provide similar insight.
The alignment of \vecJ\ with \vecB\ is nearly independent of the grain
shape, \QJ\ being very close to the ``spherical solution'' for all values of
the axis ratio (Fig.~9),
magnetic damping parameter (Fig.~10) 
and dust-to-gas temperature ratio (Fig.~11).
This is clearly a consequence of the weak dependence of the angular
momentum diffusion coefficients on the grain shape and orientation.\footnote{
This weak dependence made the perturbative approach of L95 and L97
possible.}
The monotonic increase of \QJ\ with \deltm\ (Fig.~10)
reflects the increasing
efficiency of magnetic damping as the ratio $\tgas/\tmag$ increases.
Of course, the alignment should saturate, with \QJ\ approaching some asymptotic
limit as $\deltm\rightarrow\infty$.
However, the largest \deltm\ values in Fig.~10
are too small to see this limiting behavior,
which is manifested only for the very large \deltm\ values characteristic
of superparamagnetic grains (see \S4.2).
The rapid decline of \QJ\ with increasing dust temperature (Fig.~11)
illustrates
the effects of thermal fluctuations in the magnetization, whose interaction
with \vecB\ produces random torques that tend to disorient \vecJ\ (JS67).
As $\TsTg\rightarrow 1$, a balance is reached between
the dissipation of rotational energy into heat by magnetic dissipation and the
transfer of energy to rotation by the fluctuations.
In this limit, the distribution of the transverse angular momentum components
becomes Maxwellian and the alignment of \vecJ\ vanishes.
Our solutions also satisfy this benchmark.

Our numerical code also calculates the correlations between $\theta$ and
$\beta$. Define the dimensionless ``correlation function'' by
\be
\rho \equiv
{
\left<\cos^2\theta\cos^2\beta\right>
\over
\left<\cos^2\theta\right>\,\left<\cos^2\theta\right>
}-1.
\label{eq-4_6}
\ee
In Figure~12 we plot $\rho$ vs.\ \deltm\ for the case $a/b=2/3$
and $\TsTg=0.2$.
The correlations appear to be a minor effect, in the sense that
$\left<\cos^2\theta\cos^2\beta\right>$ never differs by more than a few
percent from the product
$\left<\cos^2\theta\right>\times\left<\cos^2\theta\right>$.
However the apparent smallness of $\rho$ is misleading.
Let
\be
R \equiv \QX\QJ\left(1+\tau\right),
\label{eq-4_7}
\ee
so that $\tau=0$ corresponds to the situation where the correlations vanish.
It is easy to show from expressions (\ref{eq-2_7})--(\ref{eq-2_10}) that
$\rho$ and $\tau$ are related by
\be
\tau = {1+2Q_X+2Q_J+4Q_XQ_J\over 4 Q_XQ_J}\,\rho,
\label{eq-4_8}
\ee
so that $\tau \gg \rho$ if $\QX\QJ \ll 1$.
For example, consider the case $\deltm=1$, $\TsTg=0.2$ and $a/b=2/3$.
The alignment measures are
$\QX=0.1573$, $\QJ=8.761 \times 10^{-2}$,
$R=2.376\times 10^{-2}$
and the correlation function is $\rho=0.025$.
Thus, $R$ is a factor $\approx 1.7$ larger than the product $\QX\QJ$
even though $\rho \ll 1$.
By this measure the correlations are large.
In fact, this example is typical.
For the 239 data points in Figures~9--11, the ratio $R/\QX\QJ$ varies
from 1.3 to 5.0 with a mean value of 1.7. In contrast, 
$\rho$, never exceeds $0.1$.
We conclude that the Rayleigh reduction factor is very sensitive to
small correlations between $\theta$ and $\beta$.


Because $R$ is sensitive to the correlations, it is important
to ask whether they are real or merely an artifact of the numerics.
The benchmark calculations in Figs.~6--8 do not address
this question because they describe special cases where the 
correlations vanish.
Nevertheless, it is easy to see that the behavior of $\rho$ in Fig.~12
is qualitatively correct:
In the adiabatic approximation, the correlation function is merely a
statistic of the angular momentum distribution, \fext.
We cannot check the \deltm\ dependence of \fext\
directly because our numerical integration scheme yields only statistics
of the various distributions.
However, there can be no qualitative differences between the \deltm\ dependence
of \fext\ for grains with $a/b=2/3$ and \fext\ for spherical grains
(which can be calculated exactly, see PS71).
The dashed curve in Figure~12
is an approximation to the correlation function
obtained by replacing \fext\ with the exact solution for spheres with
$\deltm=1$ and $\TsTg=0.2$.\footnote{That is, we computed
$\left<\cos^2\theta\right>$, etc using expressions
(\ref{eq-3_5})--(\ref{eq-3_7})
for nonspherical grains with $a/b=2/3$ and $\TsTg=0.2$ and averaged
these expression
over the angular momentum distribution for spheres.}
The qualitative similarity of the two curves in Fig.~12
shows that the
correlations are not a numerical artifact.
An additional check on the correlations is discussed in \S4.2, where
we show that numerically-computed $R$ values converge to an analytic solution
for $R$ that becomes exact in the limit $\deltm\rightarrow\infty$.
Given the sensitivity of $R$ to the correlations, the convergence
can only occur if the numerical treatment of the correlations is
highly accurate.

\subsection{Superparamagnetic grains}

The magnetic susceptibility of interstellar grains can be enhanced to
``superparamagnetic'' values if the grains contain small clusters of
ferromagnetic or ferrimagnetic atoms
(JS67; Draine 1996; Draine \& Lazarian 1998, henceforth DL98).
The existence of superparamagnetic grains is consistent with the
wavelength dependence of continuum polarization
(Mathis 1986)
and is supported by the presence of small FeNi and FeNiS inclusions
in the GEMS component of interplanetary dust\footnote{DL98 claim, however,
that not more than 5\% of interstellar iron can be in the form of
metallic inclusions.} (Bradley 1994; Martin 1995;
Goodman \& Whittet 1995).
The alignment of superparamagnetic grains is therefore of considerable
interest.

Superparamagnetism corresponds to the case $1 \ll \deltm \lesssim 10^5$,
a regime where numerical calculations of the alignment are impractical.
The numerical difficulties are caused by the disparity between
the timescales for gas damping (=1 in dimensionless units) and
magnetic damping (=$\delta_{\rm m}^{-1}$).
%
(That is, the Langevin equation becomes ``stiff.'')
%
To produce accurate statistics, numerical integrations must be carried out
with a timestep much smaller than the smallest dynamical timescale and an
averaging time much larger than the largest timescale.
It follows that the number of arithmetic operations per $R$ value
increases in proportion to \deltm\ and
a numerical investigation of
superparamagnetic alignment would require orders of magnitude
more computation than the present study.

However, the disparity between \tgas\ and \tmag\ can be turned to advantage
using the adiabatic elimination
technique.
In Appendix~B we carry out the analysis to show that
\be
\fext = \fxyv\,\fzv + {\cal O}\left(\delta_{\rm m}^{-1}\right).
\label{eq-4_9}
\ee 
The angular momentum distribution functions, \fxy\ and \fz, have
closed-form solutions.
For the transverse components, we find
\be
\fxyv = C_{xy}\,\exp\left(-\Phi\right),
\label{eq-4_10}
\ee
where $C_{xy}$ (a function of \Jz) is determined by the normalization condition
\be
\int_{-\infty}^{+\infty}\,\int_{-\infty}^{+\infty}\ \fxyv\,d\Jx\,d\Jy =1.
\label{eq-4_11}
\ee
The argument of the exponential is
\be
\Phi\left(J_x,J_y|J_z\right)
\equiv \left({T_{\rm g} \over T_{\rm s}}\right)\,
\int_0^{J_{\bot}}\,
\left[1+\left(h-1\right)\sigma\left(J^{\prime}\right)\right]\,
dJ_{\bot}^{\prime},
\label{eq-4_12}
\ee
where $\Jper\equiv\left(J_x^2+J_y^2\right)^{1/2}$.
The distribution of $J_z$ is
\be
\fzv = C_z\,\Btilde\,\exp\left(-\Psi\right),
\label{eq-4_13}
\ee
where $C_z$ is a normalization constant determined by
\be
\int_{-\infty}^{+\infty}\ \fzv\,d\Jz =1
\label{eq-4_14}
\ee
and
\be
\Psi\left(J_z\right) = \int_0^{J_z}\,
\left[\,
{-2\tilde{A}_z\over\tilde{B}_{zz}}
\,\right]
dJ_z^{\prime}.
\label{eq-4_15}
\ee
The coefficients $\tilde{A}$ and $\tilde{B}$ are functions
of $J_z$ determined by averaging the diffusion coefficients
in equation (\ref{eq-3_8}) over the transverse angular
momentum components, thus:
\be
\Atildz \equiv \intf\intf \bar{A}_z\,\fxy\,dJ_x\,dJ_y
\label{eq-4_16}
\ee
and
\be
\Btildzz \equiv \intf\intf \bar{B}_{zz}\,\fxy\,dJ_x\,dJ_y.
\label{eq-4_17}
\ee
This completes the solution.

One can easily verify that solution (\ref{eq-4_9}) reduces
to the exact analytic solution for spherical grains (PS71)
in the appropriate limit.
An additional check is provided by comparing $R$ values obtained
analytically from eq.~(\ref{eq-4_9}) with numerical solutions
of the Langevin equations for large \deltm.
The results of a few tests with $\deltm=100$ are given in Table~7.
The numerical $R$ values were obtained by integrating the
Langevin equations with $\Dt=10^{-5}$ and $T=10^5$
dimensionless units.\footnote{It took 3.5~CPU days on a Sun~4 to
compute each of these points.}
According to expression (\ref{eq-4_9}),
the numerical and analytic
results should agree to within $\sim$$1$\% if the discrepancies
are due mainly to the accuracy of the adiabatic approximation
rather than errors in the numerics.
The discrepancies are $0.3$--$2$\%, consistent with this interpretation.
The fact that the numerical $R$ values are systematically
{\it smaller}\/ than
the analytic results reinforces this conclusion.

The Rayleigh reduction factor for superparamagnetic grains depends
only on $a/b$ and $\TsTg$.
Values of $R$ can be calculated straightforwardly, in principle,
from expressions (\ref{eq-4_9}) and (\ref{eq-3_11}).
However this involves the numerical evaluation of certain multiple
integrals and a large amount of computation.
We have therefore tabulated the $R$ values for superparamagnetic
grains in Table~8.

%
%

\section{Comparison with earlier work}

\subsection{Classical studies}

The objective of this paper has been to provide accurate calculations on the
Davis-Greenstein effect incorporating all physical processes now believed to
be relevant.
The important issue of how our calculations bear on the classical
``grain alignment problem'' is deferred to a future paper, where we
make a comprehensive comparison between the observational constraints,
our calculations on the DG effect and the calculations from forthcoming
papers on other alignment mechanisms.
Nevertheless, a brief comparison of the our results with previous studies of
DG alignment is appropriate.

Until recently, the standard work on DG alignment was the Monte Carlo study of PS71,
who described the alignment of oblate and prolate grains including gas-grain
collisions, evaporation of molecules from the grain surface, magnetic dissipation,
and thermal fluctuations in the grain magnetization. PS71 did not include
Barnett relaxation, whose relevance to grain alignment was not appreciated
until the work of P79.
The nature of the Monte Carlo method used by PS71 forced them to adopt some
computational expedients\footnote{For example, they modelled
the grain surface with just 4 patches, scaled up the gas/grain mass ratio
from a realistic value of $\sim 10^{-12}$ to ``something like $10^{-2}$'' and
represented the Maxwellian velocity distribution of the gas particles by two
delta functions.} but otherwise their simulations model the alignment
with complete verisimilitude.
Thus, the essential difference between our calculations and those of P71 is
the inclusion of Barnett relaxation in the present study.

In Figure~13 we compare values of \QX, \QJ\ and $R$ obtained
with our Langevin code (solid curves) and values for corresponding
cases tabulated in PS71 (symbols).
Not surprisingly, there is good agreement on \QJ,
the results from both studies being very close to the ``spherical solution''
for reasons discussed in \S4.1.
However there are large differences in the solutions for \QX.
In PS71, the motion of \veca\ in the body frame is caused by
gas damping.
Thus, the PS71 values for \QX\ are very close to
the ``Maxwellian solution''.
We assume that the motion of \veca\ is caused by Barnett relaxation.
Consequently, we find perfect alignment of \veca\ with \vecJ\ for $\TsTg=0$,
with \QX\ decreasing monotonically to the Maxwellian solution
as $\TsTg\rightarrow 1$.
The behavior of  \QX\ and \QJ\ explains why our solutions for
$R$ are systematically larger than those of PS71.\footnote{This
would not be the case if the ``correlation factor'', $\tau$, 
were different in the two studies. In fact, $\tau=0.6$--$0.7$ for the 
PS71 data plotted in Fig.~13 (see PS71, Table~2), nearly identical
to our mean $\tau$ value.}
We conclude that the Barnett effect generally increases the efficiency of
DG alignment.
The enhancement is a factor of $\approx$$3$ for $\TsTg \ll 1$ but
decreases rapidly as the dust temperature increases.
For $\TsTg \gtrsim 0.3$, there is little difference between our
solutions and those of PS71.

Figure~14 compares the \deltm\ dependence of our solutions with
analogous solutions from PS71, for the case of cold grains
with $\TsTg=1/9$.
For small \deltm, our $R$ values exceed those of PS71
by factors of 2--3.
Our results appear to converge to those of PS71 for $\deltm \gg 1$.
However, it is not clear whether the convergence occurs for
other axis- and temperature ratios, there being no obvious
physical explanation for this behavior.

\subsection{L97}


The essential difference between L97 and this paper is
that the former replaces the random variables
$\cos^2\theta$ and $\cos^2\beta$ with their mean values
in the Fokker-Planck equation.
The correlation
between $\theta$ and $\beta$, which was neglected in earlier studies,
was introduced in L97 through varying the value of $|J|$.
Therefore, it would be correct to write down the assumed dependence
of $R$ on $Q_J$ and $Q_X$
in L97 in the following way:
\begin{equation}
R\approx Q_J Q_X(J|\langle\cos^2\beta \rangle)~~~,
\end{equation}
where $J|\langle \cos^2\beta \rangle$ expresses the conditional
dependence of $J$ on $\beta$. 
We briefly discuss the perturbation
procedure in Appendix~C, as some points of it have been rectified after 
L97 was published.  In fact, this is the first
actual use of the perturbation procedure developed in L97.

To obtain $R$ we start with the Maxwellian value of angular momentum
(see LR97)
\begin{equation}
J^2_{\rm Max}=\left[1+\frac{(a/b)^2}{2}\right]\left(1+\frac{T_{\rm av}}{
T_{\rm s}}\right)
\end{equation}
corresponding
to the temperature (see L95, L97):
\begin{equation}
T_{\rm av}=\frac{T_{m}+T_{\rm s}\delta_{i-1}}{1+\delta_{i-1}}~~~.
\label{t_av}
\end{equation}
This enables us to find $\cos^2\beta$ and $\cos^2\theta$ that are used
to iterate the mean value of angular momentum and $Q_J$ and $Q_X$.
We find that the iterations quickly converge and that the mean value
of grain angular momentum is much more sensitive to variations of
$\beta$ than of $\theta$.


The angles $\beta$ and $\theta$ are correlated within the perturbative
%
approach through the value of the grain angular momentum. If we disregard
%
small terms proportional to powers of small parameter $\gamma$,
the value of the mean angular momentum that is used within the perturbative
approximation is (see (\ref{J_it}))
\begin{equation}
\langle J^2 \rangle_i \approx \frac{2kT_{\rm av}I_{\|}}{
\left(1-\cos^2\beta_i\left[1-\aleph^2_i\right]\right)
(1+\sin^2\theta_i[h-1])}
\label{v_app}
\end{equation}
where $i$ denotes the iteration number. 
It is clear  from eq.~(\ref{v_app}) that as $\cos^2 \beta$ changes
so does $\langle J^2 \rangle_i$. But according to (\ref{ddis}) 
$\langle J^2 \rangle_i$ determines the value of $Q_X$ and
therefore $\cos^2\theta$ (see eq.~(\ref{eq-2_9})). 

To calibrate the accuracy of the perturbative method, we have
carried out a point-by-point comparison between the numerical $R$
values compiled in Tables 2--6 and corresponding values obtained
with the fifth-order iterative method (see Appendix~C).
We find that the mean error in the perturbative results
is $16.8$\%, the sign indicating that the perturbative method (slightly)
overestimates the true solution. Given the various idealizations
in our grain model, this accuracy should be sufficient for many
applications.

\section{Future work}

Does our present paper mean that the theory of Davis-Greenstein alignment is
complete? We see
a couple of outstanding problems that we hope to handle in the near
future:
The present study describes oblate grains.
%
Although there exists some evidence
%
that aligned grains tend to be oblate (Hildebrand 1988) a study of prolate
grain alignment is needed.
Our solutions assume
that the magnetic properties of grains are isotropic. This is an exellent
approximation for paramagnetic grains, but ferromagnetic grains should
be anisotropic (see DL98). This means that for
sufficiently high $\delta_{\rm m}$ we have to consider the anisotropy\footnote{
In some instances, e.g., when a grain contains numerous superparamagnetic
inclusions, the magnetic response can be isotropic despite high  
$\delta_{\rm m}$. This is probably an exception, however.} of
the grain magnetic response. Therefore  a quantitative
study of anisotropy effects is necessary.

JS67 attracted the attention of researchers to the necessity of
studying the magnetic response of grains. A recent study by DL98
clarified a number of issues, especially relevant
to the high frequency magnetic response of candidate materials.
We feel that the low frequency magnetic response deserves
a similar study.
We conclude that there is plenty of work to be done on
Davis-Greenstein alignment.
We will be lucky to finish this work by the half-century anniversary of
DG51.

%
%
\section{Summary}

The principal results of this paper are as follows:

\begin{enumerate}

\item
We have calculated the efficiency of the Davis-Greenstein effect
for oblate spheroidal grains of arbitrary axis ratio.
Our description of the grain dynamics includes gas damping,
magnetic damping and Barnett relaxation.

\item
We have calculated the alignment of ordinary paramagnetic grains
in an ``adiabatic approximation'' that is accurate to
\Ordeps, where $\epsilon \sim 10^{-3}$ is the ratio of
the timescale for Barnett relaxation to the dynamical
timescale for the angular momentum.

\item
For ordinary grains, we have tabulated the Rayleigh factor, $R$, 
on a grid of $>$1000 combinations of the relevant parameters.
We have estimated the accuracy of these results using various
benchmark tests. The error in $R$ is
typically a few units in the third decimal place.

\item
The main difference between our paper and classical studies
of DG alignment is the incorporation of Barnett relaxation in the
present study.
Including the Barnett effect enhances $R$ by factors up to
$\approx 3$ for small dust-to-gas temperature ratios but
makes little difference for $\TsTg > 0.3$.


\item
We have calibrated the accuracy of an approximate method proposed in L97.
The mean error in values of $R$ computed via the
perturbative method is $\approx$$17$\%.

\item
We have derived an exact analytic solution for the alignment of
superparamagnetic grains. The solution is based on an adiabatic
approximation that is accurate to
${\cal O}\left(\delta_{\rm m}^{-1}\right)$.

\item
Our calculations on superparamagnetic grains show that DG alignment
is only efficient if the grains are much colder than the gas.
The commonly-held view that superparamagnetism produces
``perfect'' alignment
under realistic conditions is a misconception.

\end{enumerate}

\noindent
{\bf Acknowledgments} ---  We thank Bruce Draine for helpful discussions
on various aspects of grain alignment. This work was partially supported
by NASA grant NAG5-2858.

%
%

\newpage
\newcommand{\Rotwd}{\mbox{Rotation period}}          
\newcommand{\Roteo}{\mbox{$
            7 \times 10^{-5}\ 
            \rho_{{\rm s},0}^{1/2}\,
            T_{{\rm g},1}^{-1/2}\,
            (a/b)^{1/2}\,
            b_{-5}^{5/2}
            $}}
\newcommand{\Larwd}{\mbox{Larmor period}}          
\newcommand{\Lareo}{\mbox{$
            2 \times 10^5\ 
            \rho_{{\rm s},0}\,
            \left(\chi^{\prime}_{-3}\right)^{-1}\,
            B_{-5}^{-1}\,
            b_{-5}^2
            $}}
\newcommand{\Barts}{\mbox{$t_{{\rm Bar}}$\,\dotfill}}          
\newcommand{\Barwd}{\mbox{Barnett time}}
\newcommand{\Bareo}{\mbox{$
            4\times 10^6\ 
            \rho_{{\rm s},0}^2\,
            K_{-13}^{-1}\,
            T_{{\rm g},1}^{-1}\,
            {(a/b)\left(1+a^2/b^2\right)^3 \over \left(1-a^2/b^2\right)}\,
            b_{-5}^7
            $}}
\newcommand{\Gaswd}{\mbox{Gas damping time}}
\newcommand{\Gaseo}{\mbox{$7 \times 10^9\ 
            \Gamma_{\parallel}^{-1}\,
            \rho_{{\rm s},0}\,
            n_4^{-1} \,
            T_{{\rm g},1}^{-1/2}\,
            (a/b)\,
            b_{-5}
            $}}
\newcommand{\Magwd}{\mbox{Magnetic damping time}}
\newcommand{\Mageo}{\mbox{$4 \times 10^{12}\ 
            \rhosz\,
            K_{-13}^{-1}\, 
            B_{-5}^{-2}\,
            b_{-5}^2
            $}}
\begin{table}
\begin{center}
\begin{tabular}{lll}
\multicolumn{3}{c}{\bf TABLE~1 } \\
       &        &         \\     
\multicolumn{3}{c}{\bf Dynamical Timescales\tablenotemark{a}}\\
       &        &         \\     
       &        &         \\     \hline\hline
\multicolumn{1}{l}{Symbol} &
\multicolumn{1}{l}{Definition} &
\multicolumn{1}{l}{Value (s)} \\ \hline
       &        &         \\     
\trot  & \Rotwd & \Roteo  \\
       &        &         \\     
\tLar  & \Larwd & \Lareo  \\
       &        &         \\     
\tBar  & \Barwd & \Bareo  \\
       &        &         \\     
\tgas  & \Gaswd & \Gaseo  \\
       &        &         \\     
\tmag  & \Magwd & \Mageo  \\
       &        &         \\ \hline
\end{tabular}
\end{center}

\vspace{0.5in}
\tablenotetext{a}{Timescales for a homogeneous,
 oblate spheroid rotating
about its symmetry axis with kinetic energy $kT_{\rm g}/2$
%
in a  gas of pure H$_2$ (after RDGF93, Table~1).
%
The quantity $K\equiv\chipp/\omega$ depends on the uncertain grain
composition.
The scaling of $K$ is appropriate for ordinary paramagnetic substances;
for superparamagnetic grains, increase $K$ by an uncertain factor
$\lesssim 10^5$ (Draine 1996).
}
\end{table}

\newpage
\renewcommand{\arraystretch}{0.8}
\begin{table}
\begin{center}
\begin{tabular}{rccccccccc}
\multicolumn{10}{c}{\bf TABLE~2 } \\
\multicolumn{10}{c}{\bf Rayleigh Reduction factor for {\boldmath$a/b=0.1$}}\\
                                                         \\     \hline\hline
 & \multicolumn{9}{c}{$T_{\rm s}/T_{\rm g}$}\\ \cline{2-10}
\multicolumn{1}{c}{$\log\delta_{\rm m}$} &
\multicolumn{1}{c}{$0$}         &
\multicolumn{1}{c}{$.1$}         &
\multicolumn{1}{c}{$.2$}         &
\multicolumn{1}{c}{$.3$}         &
\multicolumn{1}{c}{$.4$}         &
\multicolumn{1}{c}{$.5$}         &
\multicolumn{1}{c}{$.6$}         &
\multicolumn{1}{c}{$.7$}         &
\multicolumn{1}{c}{$.8$}         \\ \hline
-1.00 & 0.018 & 0.014 & 0.008 & 0.006 & ----- & ----- & ----- & ----- & ----- \\
-0.90 & 0.022 & 0.014 & 0.010 & 0.006 & ----- & ----- & ----- & ----- & ----- \\
-0.80 & 0.030 & 0.019 & 0.011 & 0.008 & 0.005 & ----- & ----- & ----- & ----- \\
-0.70 & 0.035 & 0.023 & 0.015 & 0.009 & 0.006 & ----- & ----- & ----- & ----- \\
-0.60 & 0.046 & 0.029 & 0.018 & 0.011 & 0.007 & 0.005 & ----- & ----- & ----- \\
-0.50 & 0.055 & 0.035 & 0.021 & 0.014 & 0.009 & 0.006 & ----- & ----- & ----- \\
-0.40 & 0.066 & 0.040 & 0.025 & 0.016 & 0.011 & 0.007 & 0.005 & ----- & ----- \\
-0.30 & 0.083 & 0.049 & 0.029 & 0.019 & 0.012 & 0.009 & 0.006 & ----- & ----- \\
-0.20 & 0.101 & 0.059 & 0.034 & 0.022 & 0.014 & 0.010 & 0.006 & ----- & ----- \\
-0.10 & 0.119 & 0.068 & 0.039 & 0.025 & 0.016 & 0.011 & 0.008 & 0.005 & ----- \\
    0 & 0.142 & 0.078 & 0.045 & 0.027 & 0.019 & 0.013 & 0.008 & 0.005 & ----- \\
 0.10 & 0.169 & 0.090 & 0.050 & 0.032 & 0.021 & 0.013 & 0.009 & 0.006 & ----- \\
 0.20 & 0.200 & 0.101 & 0.056 & 0.035 & 0.022 & 0.015 & 0.010 & 0.007 & ----- \\
 0.30 & 0.228 & 0.112 & 0.061 & 0.038 & 0.025 & 0.017 & 0.011 & 0.007 & ----- \\
 0.40 & 0.263 & 0.121 & 0.066 & 0.041 & 0.026 & 0.017 & 0.012 & 0.008 & ----- \\
 0.50 & 0.300 & 0.133 & 0.070 & 0.043 & 0.028 & 0.018 & 0.013 & 0.008 & 0.005 \\
 0.60 & 0.337 & 0.140 & 0.074 & 0.045 & 0.029 & 0.020 & 0.013 & 0.008 & 0.005 \\
 0.70 & 0.372 & 0.149 & 0.078 & 0.047 & 0.030 & 0.020 & 0.013 & 0.009 & 0.005 \\
 0.80 & 0.411 & 0.154 & 0.081 & 0.050 & 0.032 & 0.021 & 0.014 & 0.009 & 0.005 \\
 0.90 & 0.452 & 0.161 & 0.082 & 0.050 & 0.032 & 0.022 & 0.014 & 0.009 & 0.005 \\
 1.00 & 0.490 & 0.167 & 0.086 & 0.051 & 0.033 & 0.022 & 0.015 & 0.009 & 0.005 \\
 1.10 & 0.527 & 0.170 & 0.087 & 0.052 & 0.034 & 0.022 & 0.015 & 0.009 & 0.005 \\
 1.20 & 0.562 & 0.175 & 0.088 & 0.053 & 0.034 & 0.022 & 0.015 & 0.009 & 0.005 \\
 1.30 & 0.597 & 0.176 & 0.089 & 0.054 & 0.034 & 0.023 & 0.015 & 0.009 & 0.005 \\
 1.40 & 0.629 & 0.179 & 0.090 & 0.054 & 0.034 & 0.023 & 0.015 & 0.009 & 0.005 \\
 1.50 & 0.661 & 0.178 & 0.089 & 0.054 & 0.034 & 0.023 & 0.014 & 0.009 & 0.005 \\ \hline
\end{tabular}
\end{center}
\end{table}

\newpage
\begin{table}
\begin{center}
\begin{tabular}{rccccccccc}
\multicolumn{10}{c}{\bf TABLE~3 } \\
\multicolumn{10}{c}{\bf Rayleigh Reduction factor for {\boldmath$a/b=0.3$}}\\
                                                         \\     \hline\hline
 & \multicolumn{9}{c}{$T_{\rm s}/T_{\rm g}$}\\ \cline{2-10}
\multicolumn{1}{c}{$\log\delta_{\rm m}$} &
\multicolumn{1}{c}{$0$}         &
\multicolumn{1}{c}{$.1$}         &
\multicolumn{1}{c}{$.2$}         &
\multicolumn{1}{c}{$.3$}         &
\multicolumn{1}{c}{$.4$}         &
\multicolumn{1}{c}{$.5$}         &
\multicolumn{1}{c}{$.6$}         &
\multicolumn{1}{c}{$.7$}         &
\multicolumn{1}{c}{$.8$}         \\ \hline
-1.00 & 0.015 & 0.011 & 0.007 & ----- & ----- & ----- & ----- & ----- & ----- \\
-0.90 & 0.021 & 0.015 & 0.009 & 0.006 & ----- & ----- & ----- & ----- & ----- \\
-0.80 & 0.026 & 0.018 & 0.011 & 0.007 & 0.005 & ----- & ----- & ----- & ----- \\
-0.70 & 0.031 & 0.021 & 0.013 & 0.008 & 0.006 & ----- & ----- & ----- & ----- \\
-0.60 & 0.042 & 0.026 & 0.016 & 0.011 & 0.007 & ----- & ----- & ----- & ----- \\
-0.50 & 0.053 & 0.033 & 0.019 & 0.012 & 0.008 & 0.005 & ----- & ----- & ----- \\
-0.40 & 0.064 & 0.039 & 0.023 & 0.015 & 0.009 & 0.006 & ----- & ----- & ----- \\
-0.30 & 0.079 & 0.046 & 0.027 & 0.018 & 0.012 & 0.008 & 0.005 & ----- & ----- \\
-0.20 & 0.098 & 0.056 & 0.032 & 0.020 & 0.013 & 0.009 & 0.006 & ----- & ----- \\
-0.10 & 0.120 & 0.064 & 0.036 & 0.023 & 0.015 & 0.010 & 0.007 & ----- & ----- \\
    0 & 0.140 & 0.074 & 0.042 & 0.026 & 0.017 & 0.011 & 0.008 & 0.005 & ----- \\
 0.10 & 0.167 & 0.083 & 0.046 & 0.028 & 0.019 & 0.012 & 0.008 & 0.006 & ----- \\
 0.20 & 0.197 & 0.094 & 0.051 & 0.032 & 0.020 & 0.014 & 0.009 & 0.005 & ----- \\
 0.30 & 0.230 & 0.102 & 0.055 & 0.034 & 0.022 & 0.015 & 0.010 & 0.006 & ----- \\
 0.40 & 0.260 & 0.113 & 0.059 & 0.036 & 0.024 & 0.015 & 0.010 & 0.007 & ----- \\
 0.50 & 0.296 & 0.120 & 0.064 & 0.039 & 0.025 & 0.016 & 0.011 & 0.007 & ----- \\
 0.60 & 0.336 & 0.128 & 0.066 & 0.040 & 0.026 & 0.018 & 0.012 & 0.007 & ----- \\
 0.70 & 0.375 & 0.136 & 0.070 & 0.042 & 0.028 & 0.018 & 0.012 & 0.008 & ----- \\
 0.80 & 0.412 & 0.141 & 0.073 & 0.043 & 0.028 & 0.019 & 0.013 & 0.007 & 0.005 \\
 0.90 & 0.451 & 0.148 & 0.074 & 0.044 & 0.029 & 0.019 & 0.013 & 0.008 & 0.005 \\
 1.00 & 0.490 & 0.151 & 0.076 & 0.046 & 0.029 & 0.020 & 0.013 & 0.008 & 0.005 \\
 1.10 & 0.526 & 0.154 & 0.078 & 0.046 & 0.030 & 0.020 & 0.013 & 0.008 & 0.005 \\
 1.20 & 0.562 & 0.155 & 0.077 & 0.046 & 0.029 & 0.020 & 0.013 & 0.008 & 0.005 \\
 1.30 & 0.600 & 0.159 & 0.080 & 0.047 & 0.030 & 0.020 & 0.013 & 0.008 & 0.005 \\
 1.40 & 0.629 & 0.161 & 0.079 & 0.047 & 0.031 & 0.020 & 0.013 & 0.008 & 0.005 \\
 1.50 & 0.663 & 0.161 & 0.081 & 0.047 & 0.030 & 0.020 & 0.013 & 0.008 & ----- \\ \hline
\end{tabular}
\end{center}
\end{table}

\newpage
\begin{table}
\begin{center}
\begin{tabular}{rcccccccc}
\multicolumn{9}{c}{\bf TABLE~4 } \\
\multicolumn{9}{c}{\bf Rayleigh Reduction factor for {\boldmath$a/b=0.5$}}\\
 & \multicolumn{8}{c}{$T_{\rm s}/T_{\rm g}$}\\ \cline{2-9}
\multicolumn{1}{c}{$\log\delta_{\rm m}$} &
\multicolumn{1}{c}{$0$}         &
\multicolumn{1}{c}{$.1$}         &
\multicolumn{1}{c}{$.2$}         &
\multicolumn{1}{c}{$.3$}         &
\multicolumn{1}{c}{$.4$}         &
\multicolumn{1}{c}{$.5$}         &
\multicolumn{1}{c}{$.6$}         &
\multicolumn{1}{c}{$.7$}         \\ \hline
-1.00 & 0.014 & 0.010 & 0.005 & ----- & ----- & ----- & ----- & ----- \\
-0.90 & 0.017 & 0.012 & 0.007 & 0.005 & ----- & ----- & ----- & ----- \\
-0.80 & 0.022 & 0.015 & 0.009 & 0.005 & ----- & ----- & ----- & ----- \\
-0.70 & 0.032 & 0.020 & 0.011 & 0.007 & 0.005 & ----- & ----- & ----- \\
-0.60 & 0.039 & 0.024 & 0.013 & 0.008 & 0.006 & ----- & ----- & ----- \\
-0.50 & 0.049 & 0.029 & 0.016 & 0.010 & 0.007 & 0.005 & ----- & ----- \\
-0.40 & 0.064 & 0.035 & 0.019 & 0.012 & 0.008 & 0.005 & ----- & ----- \\
-0.30 & 0.079 & 0.041 & 0.022 & 0.014 & 0.009 & 0.006 & ----- & ----- \\
-0.20 & 0.095 & 0.048 & 0.026 & 0.016 & 0.010 & 0.007 & 0.005 & ----- \\
-0.10 & 0.115 & 0.056 & 0.031 & 0.018 & 0.012 & 0.008 & 0.005 & ----- \\
    0 & 0.139 & 0.064 & 0.034 & 0.021 & 0.013 & 0.009 & 0.006 & ----- \\
 0.10 & 0.167 & 0.072 & 0.037 & 0.023 & 0.015 & 0.010 & 0.006 & ----- \\
 0.20 & 0.195 & 0.079 & 0.041 & 0.025 & 0.016 & 0.011 & 0.007 & ----- \\
 0.30 & 0.228 & 0.088 & 0.044 & 0.026 & 0.017 & 0.011 & 0.008 & 0.005 \\
 0.40 & 0.262 & 0.096 & 0.047 & 0.028 & 0.018 & 0.012 & 0.008 & 0.005 \\
 0.50 & 0.298 & 0.101 & 0.051 & 0.030 & 0.019 & 0.013 & 0.009 & 0.006 \\
 0.60 & 0.335 & 0.105 & 0.052 & 0.032 & 0.021 & 0.013 & 0.009 & 0.006 \\
 0.70 & 0.371 & 0.111 & 0.055 & 0.032 & 0.020 & 0.014 & 0.009 & 0.006 \\
 0.80 & 0.409 & 0.116 & 0.056 & 0.034 & 0.022 & 0.015 & 0.009 & 0.006 \\
 0.90 & 0.453 & 0.120 & 0.059 & 0.034 & 0.022 & 0.015 & 0.010 & 0.006 \\
 1.00 & 0.492 & 0.124 & 0.060 & 0.035 & 0.022 & 0.015 & 0.010 & 0.006 \\
 1.10 & 0.527 & 0.126 & 0.061 & 0.035 & 0.023 & 0.015 & 0.010 & 0.006 \\
 1.20 & 0.564 & 0.127 & 0.062 & 0.036 & 0.023 & 0.015 & 0.010 & 0.006 \\
 1.30 & 0.598 & 0.129 & 0.062 & 0.037 & 0.023 & 0.015 & 0.010 & 0.006 \\
 1.40 & 0.631 & 0.131 & 0.062 & 0.036 & 0.023 & 0.016 & 0.010 & 0.006 \\
 1.50 & 0.663 & 0.131 & 0.063 & 0.037 & 0.024 & 0.015 & 0.010 & 0.006 \\ \hline
\end{tabular}
\end{center}
\end{table}

\newpage
\begin{table}
\begin{center}
\begin{tabular}{rccccccc}
\multicolumn{8}{c}{\bf TABLE~5} \\
\multicolumn{8}{c}{\bf Rayleigh Reduction factor for {\boldmath$a/b=0.7$}}\\
                                                         \\     \hline\hline
 & \multicolumn{7}{c}{$T_{\rm s}/T_{\rm g}$}\\ \cline{2-8}
\multicolumn{1}{c}{$\log\delta_{\rm m}$} &
\multicolumn{1}{c}{$0$}         &
\multicolumn{1}{c}{$.1$}         &
\multicolumn{1}{c}{$.2$}         &
\multicolumn{1}{c}{$.3$}         &
\multicolumn{1}{c}{$.4$}         &
\multicolumn{1}{c}{$.5$}         &
\multicolumn{1}{c}{$.6$}         \\ \hline
-1.00 & 0.013 & 0.007 & ----- & ----- & ----- & ----- & ----- \\
-0.90 & 0.020 & 0.008 & 0.005 & ----- & ----- & ----- & ----- \\
-0.80 & 0.027 & 0.012 & 0.006 & ----- & ----- & ----- & ----- \\
-0.70 & 0.032 & 0.015 & 0.007 & ----- & ----- & ----- & ----- \\
-0.60 & 0.042 & 0.018 & 0.009 & 0.005 & ----- & ----- & ----- \\
-0.50 & 0.053 & 0.020 & 0.011 & 0.006 & ----- & ----- & ----- \\
-0.40 & 0.064 & 0.025 & 0.012 & 0.007 & 0.005 & ----- & ----- \\
-0.30 & 0.080 & 0.030 & 0.015 & 0.008 & 0.005 & ----- & ----- \\
-0.20 & 0.096 & 0.035 & 0.017 & 0.010 & 0.006 & ----- & ----- \\
-0.10 & 0.118 & 0.039 & 0.020 & 0.012 & 0.007 & 0.005 & ----- \\
    0 & 0.141 & 0.046 & 0.021 & 0.012 & 0.008 & 0.005 & ----- \\
 0.10 & 0.167 & 0.050 & 0.024 & 0.014 & 0.009 & 0.006 & ----- \\
 0.20 & 0.195 & 0.056 & 0.027 & 0.015 & 0.010 & 0.007 & 0.005 \\
 0.30 & 0.227 & 0.063 & 0.029 & 0.017 & 0.011 & 0.007 & 0.005 \\
 0.40 & 0.261 & 0.066 & 0.030 & 0.018 & 0.011 & 0.008 & 0.005 \\
 0.50 & 0.296 & 0.071 & 0.032 & 0.019 & 0.012 & 0.008 & 0.005 \\
 0.60 & 0.335 & 0.073 & 0.033 & 0.020 & 0.013 & 0.008 & 0.006 \\
 0.70 & 0.376 & 0.077 & 0.035 & 0.020 & 0.013 & 0.009 & 0.006 \\
 0.80 & 0.413 & 0.079 & 0.036 & 0.021 & 0.013 & 0.009 & 0.006 \\
 0.90 & 0.451 & 0.082 & 0.037 & 0.021 & 0.014 & 0.009 & 0.006 \\
 1.00 & 0.490 & 0.083 & 0.038 & 0.022 & 0.014 & 0.009 & 0.006 \\
 1.10 & 0.528 & 0.084 & 0.039 & 0.022 & 0.014 & 0.009 & 0.006 \\
 1.20 & 0.563 & 0.085 & 0.039 & 0.022 & 0.015 & 0.009 & 0.006 \\
 1.30 & 0.598 & 0.087 & 0.039 & 0.023 & 0.015 & 0.010 & 0.006 \\
 1.40 & 0.630 & 0.088 & 0.040 & 0.023 & 0.015 & 0.010 & 0.006 \\
 1.50 & 0.663 & 0.088 & 0.040 & 0.023 & 0.015 & 0.010 & 0.006 \\ \hline
\end{tabular}
\end{center}
\end{table}

\newpage
\begin{table}
\begin{center}
\begin{tabular}{rccccc}
\multicolumn{6}{c}{\bf TABLE~6} \\
\multicolumn{6}{c}{\bf Rayleigh reduction factor for {\boldmath$a/b=0.9$}}\\
                                                         \\     \hline\hline
 & \multicolumn{5}{c}{$T_{\rm s}/T_{\rm g}$}\\ \cline{2-6}
\multicolumn{1}{c}{$\log\delta_{\rm m}$} &
\multicolumn{1}{c}{$0$}         &
\multicolumn{1}{c}{$.1$}         &
\multicolumn{1}{c}{$.2$}         &
\multicolumn{1}{c}{$.3$}         &
\multicolumn{1}{c}{$.4$}         \\ \hline
-1.00 & 0.015 & ----- & ----- & ----- & ----- \\
-0.90 & 0.021 & ----- & ----- & ----- & ----- \\
-0.80 & 0.026 & ----- & ----- & ----- & ----- \\
-0.70 & 0.035 & 0.005 & ----- & ----- & ----- \\
-0.60 & 0.042 & 0.007 & ----- & ----- & ----- \\
-0.50 & 0.056 & 0.008 & ----- & ----- & ----- \\
-0.40 & 0.068 & 0.009 & ----- & ----- & ----- \\
-0.30 & 0.082 & 0.011 & 0.005 & ----- & ----- \\
-0.20 & 0.100 & 0.013 & 0.006 & ----- & ----- \\
-0.10 & 0.119 & 0.015 & 0.006 & ----- & ----- \\
    0 & 0.141 & 0.016 & 0.007 & ----- & ----- \\
 0.10 & 0.169 & 0.018 & 0.008 & 0.005 & ----- \\
 0.20 & 0.198 & 0.020 & 0.009 & 0.005 & ----- \\
 0.30 & 0.227 & 0.022 & 0.010 & 0.005 & ----- \\
 0.40 & 0.261 & 0.023 & 0.010 & 0.006 & ----- \\
 0.50 & 0.300 & 0.024 & 0.011 & 0.006 & ----- \\
 0.60 & 0.335 & 0.025 & 0.011 & 0.006 & ----- \\
 0.70 & 0.372 & 0.026 & 0.012 & 0.007 & ----- \\
 0.80 & 0.414 & 0.027 & 0.012 & 0.007 & ----- \\
 0.90 & 0.450 & 0.028 & 0.012 & 0.007 & 0.005 \\
 1.00 & 0.489 & 0.029 & 0.013 & 0.007 & 0.005 \\
 1.10 & 0.526 & 0.029 & 0.013 & 0.007 & 0.005 \\
 1.20 & 0.563 & 0.030 & 0.013 & 0.007 & 0.005 \\
 1.30 & 0.598 & 0.030 & 0.013 & 0.007 & 0.005 \\
 1.40 & 0.631 & 0.030 & 0.013 & 0.008 & 0.005 \\
 1.50 & 0.661 & 0.030 & 0.013 & 0.008 & 0.005 \\ \hline
\end{tabular}
\end{center}
\end{table}

\clearpage
\centerline{\bf TABLE~7}
\centerline{\bf Analytic Solution for Superparamagnetic Grains}
\centerline{\bf vs.\ Numerical Solution for \boldmath$\delta_{\rm m}=100$}

\bigskip
\begin{center}
\begin{tabular}{cccc} \hline\hline
\multicolumn{1}{c}{$a/b$}&
\multicolumn{1}{c}{$T_{\rm s}/T_{\rm g}$}&
\multicolumn{1}{c}{Analytic}&
\multicolumn{1}{c}{Numerical}\\ \hline
0.1 & 0.1 & 0.1887 & 0.1857 \\
0.1 & 0.2 & 0.0951 & 0.0940 \\
0.1 & 0.3 & 0.0571 & 0.0559 \\
0.1 & 0.4 & 0.0370 & 0.0366 \\
0.5 & 0.1 & 0.1369 & 0.1353 \\
0.5 & 0.3 & 0.0388 & 0.0380 \\
0.5 & 0.5 & 0.0166 & 0.0165 \\ \hline
\end{tabular}
\end{center}


\bigskip
\bigskip
\bigskip
\centerline{\bf TABLE~8}
\centerline{\bf Rayleigh reduction factor for superparamagnetic grains}

\begin{center}
\begin{tabular}{cccccccccc}
\\     \hline\hline
 & \multicolumn{9}{c}{$T_{\rm s}/T_{\rm g}$}\\ \cline{2-10}
\multicolumn{1}{c}{$a/b$}        &
\multicolumn{1}{c}{$.1$}         &
\multicolumn{1}{c}{$.2$}         &
\multicolumn{1}{c}{$.3$}         &
\multicolumn{1}{c}{$.4$}         &
\multicolumn{1}{c}{$.5$}         &
\multicolumn{1}{c}{$.6$}         &
\multicolumn{1}{c}{$.7$}         &
\multicolumn{1}{c}{$.8$}         &
\multicolumn{1}{c}{$.9$}         \\ \hline
  0.1   & .1887 & .0951 & .0571 & .0370 & .0247 & .0164 & .0106 & .0061 & .0027  \\ 
  0.3   & .1688 & .0839 & .0502 & .0324 & .0217 & .0144 & .0093 & .0054 & .0024  \\ 
  0.5   & .1369 & .0657 & .0388 & .0250 & .0166 & .0111 & .0071 & .0041 & .0018  \\ 
  0.7   & .0916 & .0417 & .0242 & .0155 & .0103 & .0068 & .0044 & .0025 & .0011  \\ 
  0.9   & .0316 & .0138 & .0080 & .0051 & .0034 & .0023 & .0014 & .0008 & .0004  \\ \hline 
\end{tabular}
\end{center}

%
%

%
%
\clearpage
\appendix

\section{Diffusion coefficients in equation (3-1)}

\subsection{Gas damping}

The diffusion coefficients for gas damping were derived elsewhere
(LR97; L97).
The (dimensionless) mean torque has body frame components
\be
A_{{\rm g},x}^{\rm b} = -h\left(\Grat\right)\,J^{\rm b}_x,
\label{eq-A_1}
\ee
\be
A_{{\rm g},y}^{\rm b} = -h\left(\Grat\right)\,J^{\rm b}_y,
\label{eq-A_2}
\ee
and
\be
A_{{\rm g},z}^{\rm b} = -J^{\rm b}_z.
\label{eq-A_3}
\ee
The diffusion tensor is diagonal in the body frame with 
\be
B_{{\rm g},xx}^{\rm b} = \left(\Grat\right)\,
    \left(1+T_{\rm s}/T_{\rm g}\right),
\label{eq-A_4}
\ee
\be
B_{{\rm g},yy}^{\rm b} = B_{{\rm g},xx}^{\rm b},
\label{eq-A_5}
\ee
and
\be
B_{{\rm g},zz}^{\rm b} = \left(1+T_{\rm s}/T_{\rm g}\right).
\label{eq-A_6}
\ee
The dimensionless shape factors
\be
\Gamma_{\parallel}(e) =
{3 \over 16} \, \left\{\
3+4(1-e^2)g(e)-e^{-2}\left[1-(1-e^2)^2g(e)\right]
\right\}
\label{eq-A_7}
\ee
and
\be
\Gamma_{\perp}(e) =
{3 \over 32} \, \left\{\
7-e^2+(1-e^2)^2g(e)+
(1-2e^2)\left[1+e^{-2}\left[1-(1-e^2)^2g(e)\right]\right]\right\},
\label{eq-A_8}
\ee
are weak functions of the eccentricity, $e\equiv\sqrt{1-a^2/b^2}$,
with
\be
g(e) \equiv {1 \over 2e} \ln\left({1+e \over 1-e}\right).
\label{eq-A_9}
\ee

The diffusion coefficients in equation (\ref{eq-3_1}) are
evaluated in the inertial frame.
To obtain these coefficients from expressions
(\ref{eq-A_1})--(\ref{eq-A_6}) we
(i) transform components from the body frame to the inertial frame; and
(ii) average the resulting expressions over the rapidly-changing
angles $\phi$ and $\psi$ (see L97).
The calculations are straightforward but tedious and here
we simply give the results.
The mean torque has inertial frame components
\be
A_{{\rm g},i} = -\left(\,1+\lambda\,\sin^2\theta\,\right)J_i
\ \ \ i=x,y,z,
\label{eq-A_10}
\ee
where
\be
\lambda \equiv h\left(\Grat\right)-1.
\label{eq-A_11}
\ee
The diffusion tensor has inertial frame components
\be
B_{{\rm g},xx} = \frac{1}{2}\left(1+\TsTg\right)
\left[\,\left(1+\cos^2\beta\right)\,P_1+\sin^2\beta\,P_2\,\right],
\label{eq-A_12}
\ee
\be
B_{{\rm g},yy} = B_{{\rm g},xx},
\label{eq-A_13}
\ee
and
\be
B_{{\rm g},zz} =
\left(1+\TsTg\right)
\left[\,\sin^2\beta\,P_1+\cos^2\beta\,P_2\,\right],
\label{eq-A_14}
\ee
where
\be
P_1 \equiv 1-\gamma\left(1-\sin^2\theta/2\right)
\label{eq-A_15}
\ee
and
\be
P_2 \equiv \left(1-\gamma\sin^2\theta\right)
\label{eq-A_16}
\ee
are functions of $\theta$ and
\be
\gamma \equiv 1-\Gper/\Gpar.
\label{eq-A_17}
\ee

\subsection{Magnetic damping}

The mean torque was derived by DG51 (see DG51, eq.\ [81]).
In dimensionless units, the inertial frame components are
\be
A_{{\rm m},x} = -\Zthet\,\deltm\,J_x,
\label{eq-A_18}
\ee
\be
A_{{\rm m},y} = -\Zthet\,\deltm\,J_y,
\label{eq-A_19}
\ee
and
\be
A_{{\rm m},z} = 0,
\label{eq-A_20}
\ee
where
\be
\Zthet \equiv \Zfac.
\label{eq-A_21}
\ee

The diffusion tensor for paramagnetic
relaxation in spheroidal grains does not appear in previous
papers on the DG effect, which either ignored 
thermal fluctuations (e.g., DG51) or used the
diffusion tensor for spheres (JS67).
However the diffusion tensor
is uniquely determined by the principle of detailed balance,
which requires the probability current at each point in phase space
to vanish in thermodynamic equilibrium.
Setting the current along the $x$ direction to
zero yields
\be
A_{{\rm m},x}\,f^* -
\frac{1}{2}{\partial\over\partial J_x}\left(B_{{\rm m},xx}\,f^*\right) =0,
\label{eq-A_22}
\ee
where
\be
f^* = C\,\exp\left(\, {ZJ^2 \over 2 T_{\rm s}/T_{\rm g}} \,\right)
\label{eq-A_23}
\ee
is the thermal equilibrium distribution of the dimensionless
angular momentum and $C$ is a normalization constant.
Consistent with the symmetry of the problem, we have
assumed that the off-diagonal components of the diffusion tensor are zero.
The unique solution of eq.~(\ref{eq-A_23}) with
$B_{{\rm m},xx}$ finite for all $J_x$ is
\be
B_{{\rm m},xx} = 2\left(\TsTg\right)\deltm.
\label{eq-A_24}
\ee
The analogous conditions for the $y$ and $z$ directions yield
\be
B_{{\rm m},yy} = B_{{\rm m},xx}
\label{eq-A_25}
\ee
and
\be
B_{{\rm m},zz} = 0.
\label{eq-A_26}
\ee

%
%
\clearpage

\section{Analytic solution for superparamagnetic grains}

\subsection{Adiabatic elimination}

We seek an approximate solution to equation (\ref{eq-3_8}) of the form
\be
\fextv = \fxyv\,\fzv
\label{eq-B_1}
\ee 
subject to the normalization conditions
\be
\int_{-\infty}^{+\infty}\,\int_{-\infty}^{+\infty}\ \fxyv\,d\Jx\,d\Jy =1
\label{eq-B_2}
\ee
and
\be
\int_{-\infty}^{+\infty}\ \fzv\,d\Jz =1.
\label{eq-B_3}
\ee
We require our approximation to be accurate to \OrdmI.
This means that we may neglect the terms of order unity
in comparison with the terms of order \deltm\ in 
expressions (\ref{eq-3_12})--(\ref{eq-3_14}) and
(\ref{eq-3_19})--(\ref{eq-3_21}) for the diffusion coefficients.
For the linear coefficients, this gives
\be
\Abarx \approx -\left[1+(h-1)\sigma\right]\,J_x\,\deltm
        \equiv \Atildx\,\deltm,
\label{eq-B_4}
\ee
\be
\Abary \approx -\left[1+(h-1)\sigma\right]\,J_y\,\deltm
        \equiv \Atildy\,\deltm,
\label{eq-B_5}
\ee
and
\be
\Abarz = -\left(1+\lambda\sigma\right)\,J_z.
\label{eq-B_6}
\ee
The analogous approximations for the quadratic diffusion coefficients
are
\be
\Bbarxx \approx 2\left(\Ts/\Tg\right)\,\deltm \equiv \Btildxx\,\deltm,
\label{eq-B_7}
\ee
\be
\Bbaryy \approx \Btildxx\,\deltm,
\label{eq-B_8}
\ee
and
\be
\Bbarzz = \left(1+\TsTg\right)
\left[\,\sin^2\beta\,P_1(J)+\cos^2\beta\,P_2(J)\,\right]~~~,
\label{eq-B_9}
\ee
where $P_1(J)$ and $P_2(J)$ are defined in eqs.~(\ref{eq-A_15}) and
(\ref{eq-A_16}).
The physical content of our approximation is that the torques
perpendicular to \vecB\ are provided entirely by magnetic dissipation;
evidently this is accurate to order $\tmag/\tgas$.

To proceed, we substitute expressions (\ref{eq-B_1}),
(\ref{eq-B_4})--(\ref{eq-B_6}) and (\ref{eq-B_7})--(\ref{eq-B_9})
into equation (\ref{eq-3_8}).
After a little algebra, the latter becomes
\be
\fz\Lxy\fxy+\deltmI\Lz\fxy\fz =0,
\label{eq-B_10}
\ee
where
\be
\Lxy \equiv 
-{\partial\over\partial J_x}\,\Atildx\,\vdot
-{\partial\over\partial J_y}\,\Atildy\,\vdot
+\frac{1}{2}{\partial^2\over\partial J_x^2}\,\Btildxx\,\vdot
+\frac{1}{2}{\partial^2\over\partial J_y^2}\,\Btildxx\,\vdot
\label{eq-B_11}
\ee
and
\be
\Lz \equiv
-{\partial\over\partial J_z}\,\Abarz\,\vdot
+\frac{1}{2}{\partial^2\over\partial J_z^2}\,\Bbarzz\,\vdot.
\label{eq-B_12}
\ee

\subsection{Solution for \boldmath$f_{xy}$}

Neglecting the term of order \deltmI\ in equation (\ref{eq-B_10}) gives
\be
\Lxy\fxy=0,
\label{eq-B_13}
\ee
which is a Fokker-Planck equation for \fxy.
After transforming\footnote{The transformation is straightforward
but tedious and only the relevant results are given here. For
details, see any text on statistical mechanics  (e.g., Risken 1984).}
to polar coordinates 
\be
\Jper \equiv \left(J_x^2+J_y^2\right)^{1/2}
\label{eq-B_14}
\ee
and
\be
\phi \equiv \tan^{-1}\left(J_y/J_x\right),
\label{eq-B_15}
\ee
equation (\ref{eq-B_13}) becomes
\be
-{\partial\over\partial J_{\bot}}
\left[\,
\Aper\fper-\frac{1}{2}\Btildxx{\partial f_{\bot}\over\partial J_{\bot}}
\,\right]
-
{\partial\over\partial \phi}
\left[\,
-\frac{1}{2}\left(\Btildxx/J_{\bot}^2\right){\partial f_{\bot}\over\partial\phi}
\,\right]
=0,
\label{eq-B_16}
\ee
where
\be
\fper \equiv\Jper\fxy
\label{eq-B_17}
\ee
and
\be
\Aper \equiv -\left[1+(h-1)\sigma\right]\Jper + {\tilde{B}_{xx}\over 2J_{\bot}}.
\label{eq-B_18}
\ee
Equation (\ref{eq-B_16}) will be satisfied if we assume that \fper\ is
independent of $\phi$, consistent with the symmetry of the
problem, and require
\be
\Aper\fper-\frac{1}{2}\Btildxx{d f_{\bot}\over dJ_{\bot}}=0.
\label{eq-B_19}
\ee
Equation (\ref{eq-B_19}) states that there is no probability current
perpendicular to the \vecB\ direction in equilibrium
(``zero-current solution'' for \fper).

Solving eq.~(\ref{eq-B_19}) for \fper\ and using eq.~(\ref{eq-B_17}),
we find
\be
\fxy = C_x\,\exp\left(-\Phi\right),
\label{eq-B_20}
\ee
where $C_x$ is determined by eq.\ (\ref{eq-B_2})
and
\be
\Phi\left(J_x,J_y|J_z\right)
\equiv \left({T_{\rm g} \over T_{\rm s}}\right)\,
\int_0^{J_{\bot}}\,
\left[1+\left(h-1\right)\sigma\left(J^{\prime}\right)\right]\,
dJ_{\bot}^{\prime}.
\label{eq-B_21}
\ee
Since \Lxy\ does not involve differentiation with respect to
$J_z$, the latter appears\footnote{Recall that $\sigma$ depends
on $J^2$, hence $J_z$.} only as a parameter in the solution for \fxy.
In physical terms, this means that the distribution of $J_x$
and $J_y$ relaxes rapidly (``adiabatically'') to the equilibrium
distribution that would obtain if $J_z$ were frozen at its
instantaneous value.
Of course, this is appropriate when $\tmag\ll\tgas$.

\subsection{Solution for \boldmath$f_z$}

Given the solution for \fxy\ from the preceding section,
one can can find \fz\ by setting the term \OrdmI\ in equation
(\ref{eq-B_10}) to zero.
When eq.\ (\ref{eq-B_12}) is used to expand \Lz\ in terms of derivatives,
this statement becomes
\be
-{\partial\over\partial J_z}\left[\,\Abarz\fxy\fz\,\right]
+
\frac{1}{2}{\partial^2\over\partial J_z^2}\left[\,\Bbarzz\fxy\fz\,\right]
=0.
\label{eq-B_22}
\ee
If the last equation is multiplied by $dJ_xdJ_y$ and the
resulting expression is integrated over all $J_x$ and $J_y$,
the result is
\be
-{d\over dJ_z}\left[\,
\Atildz\fz-\frac{1}{2}{d\over dJ_z}\left(\Btildzz\fz\right)\,\right]
=0,
\label{eq-B_23}
\ee
where
\be
\Atildz \equiv \intf\intf\Abarz\fxy\,dJ_x\,dJ_y,
\label{eq-B_24}
\ee
and
\be
\Btildzz \equiv \intf\intf\Bbarzz\fxy\,dJ_x\,dJ_y.
\label{eq-B_25}
\ee
Equations (\ref{eq-B_23})--(\ref{eq-B_25}) say that \fz\ satisfies
a one-variable Fokker-Planck equation wherein the diffusion
coefficients \Abar\ and \Bbar\ have been averaged over
$J_x$ and $J_y$ in the obvious way.
The preceding discussion shows that this
procedure is accurate to \OrdmI.

The zero-current solution for \fz\ is elementary:
\be
\fz = C_z\,\tilde{B}_{zz}^{-1}\,\exp\left(-\Psi\right),
\label{eq-B_26}
\ee
where
\be
\Psi \equiv \int_0^{J_z}\,
\left({-2\tilde{A}_z\over\tilde{B}_{zz}}\right)\,dJ_z^{\prime}.
\label{eq-B_27}
\ee
This completes the solution.

\section{Perturbative scheme}

It was shown in L97 (see also Errata in Lazarian 1998) that 
calculations of $\langle Q_J \rangle$ can be performed perturbatively
with 
\begin{equation}
 \langle Q_{Ji} \rangle\approx \frac{3}{2}\langle\cos^2\beta_{i}\rangle-\frac{1}{2}~~~,
\end{equation}
where 
\begin{equation}
\langle \cos^{2}\beta_{i+1}\rangle=\frac{1}{1-\aleph_{i+1}^{2}}
\left[1-\frac{\aleph_{i+1}}{\sqrt{1-\aleph_{i+1}^{2}}}
{\rm arcsin}\sqrt{1-\aleph_{i+1}^{2}}\right]~~~,
\label{d42}
\end{equation}
\begin{equation}
\aleph_i=\frac{T_{\rm av}}{T_{\rm m}}+\frac{\gamma}{1+\delta_{\rm m}}
Q_{Xi}Q_{Ji}~~~,
\end{equation}
$T_{\rm av}$ is given by (\ref{t_av}) and $T_{\rm m}$ is the mean rotational
temperature of grain in the absence of paramagnetic relaxation. For
the sake of simplicity this temperature is frequently chosen to
be equal $0.5(T_{\rm s}+T_{\rm g})$, but such a choice is uncertain
(see Purcell 1979).

The internal alignment measure $Q_Xi$ can be found as (LR97, L97)
\begin{equation}
Q_{Xi}=\frac{3}{2}\left(
\frac{\int^{\pi}_{0}\cos^2\theta \sin\theta f(\theta)
{\rm d}\theta}{\int^{\pi}_{0}\sin\theta f(\theta){\rm d}\theta}-\frac{1}{3}
\right)~~~,
\label{ddis}
\end{equation}
\begin{equation}
f(\theta)={\rm const} \times 
\sin\theta \exp\left[-E_{\rm rot}^{(i)}(\theta)/kT_s\right]~~~,
\label{f_beta}
\end{equation}
and 
\begin{equation}
E_{\rm rot}^{(i)}(\theta)=
\frac{\langle J^2\rangle_i}{2I_z}\left[1+(h-1)\sin^2\theta\right]~~~,
\label{e_beta}
\end{equation}
where $\langle J^2 \rangle_i$ is to be found using the distribution
function  for the $i$-th iteration is
\begin{equation}
f=f_{x_0}f_{y_0}f_{z_0} \approx  {\rm const}\times
  \exp(-\alpha J^2)
\label{distri}
\end{equation}
where
  
\begin{equation}
\alpha=\frac{(1+\sin^2\theta_i[h\Gamma_{\bot}/\Gamma_{\|}-1])}{2kT_{\rm av}I_{\|}}
              \frac{J^2\left(1-\cos^2\beta_i\left[1-\aleph^2_i\right]\right)}
                   1-\gamma
                          \frac{\eta_2(\beta_i, \theta_i)T_{\rm m}}
                               {T_{\rm m}+T_{\rm s}\delta_{\rm m}}~~~,
\end{equation}
\begin{equation}
\eta_2(\beta_i,\theta_i)=\frac{1}{2}\left[\frac{1}{2}(1+\cos^2\theta)(1+\cos^2\beta)+
\sin^2\beta\sin^2\theta\right]~~~~.
\end{equation}

In the perturbative approach all the values obtained at the $i$ stage
are treated as constants and therefore integration provides
\begin{equation}
\langle J \rangle_i=const\times \int_0^{\infty}J^2 \exp(\alpha J) JdJ=
\frac{1}{2\alpha}
\label{J_it}
\end{equation}

%
%
\clearpage
\centerline{\bf FIGURE CAPTIONS}

\noindent
{\bf Fig.~1} ---
Definition of the body frame, $\left\{\bfbasis\right\}$, with
\bfzh\ parallel
to the symmetry axis, \veca, and the remaining basis vectors
oriented arbitrarily in the equatorial plane of the grain surface
(the lightly-shaded disk).
The orientation of \veca\ with respect to the angular momentum,
\vecJ, is specified by angles $\theta$ and $\psi$.
The ``$J$ frame'' basis, $\left\{\Jfbasis\right\}$, is used only to carry out
transformations between the body and inertial frames (see Fig.~2).

\noindent
{\bf Fig.~2} ---
Definition of the inertial frame, $\left\{\gfbasis\right\}$, with
\gfzh\ parallel
to the interstellar magnetic field, \vecB, and the other basis
vectors oriented arbitrarily in the plane perpendicular to \vecB.
Angles $\beta$ and $\phi$ specify
the orientation of \vecJ\ with respect to \vecB.

\bigskip
\noindent
{\bf Fig.~3} ---
The cross sections for polarized light are defined with
respect to the observer frame basis
[see eqs.\ (\ref{eq-2_5}) and (\ref{eq-2_6})],
with \ofzh\ the direction of propagation and \ofyh\ along
the projection of \vecB\ onto the plane of the sky.

\noindent
{\bf Fig.~4} ---
In the adiabatic approximation, statistics of the $\theta$ distribution
depend only on $\xi^2$ [a dimensionless function of $J^2$,
see eq.~(\ref{eq-3_4})], the grain
shape and the dust-to-gas temperature ratio.
Shown here are the mean values of $\cos^2\theta$ (short dash),
$\sin^2\theta$ (long dash), and \qX\ (solid).

\noindent
{\bf Fig.~5} ---
The angular momentum diffusion coefficients, and hence the Rayleigh
reduction factor, depend on the grain shape.
For a homogeneous grain, the dependence is only via the functions
of $a/b$ shown here:
$\lambda$ (short dash), $\gamma$ (long dash), and $h-1$ (solid).

\noindent
{\bf Fig.~6} --- Histogram: distribution of the errors in $Q_J$, computed
by comparing the results of 100 identical numerical trials
with exact solution for spherical grains (see \S4.1).
The mean and rms errors in \QJ\ are $-1.5\times 10^{-4}$
and $1.2 \times 10^{-3}$, respectively.
Dashed curve: Gaussian distribution with the same mean and standard
deviation as the histogram.

\noindent
{\bf Fig.~7} --- Comparison between \QJ\ values computed
numerically (open circles) and the predictions of the
perturbative approach described in Appendix~C.
The calculations shown here are for the case $\TsTg=0$,
$a/b=2/3$.

\noindent
{\bf Fig.~8} --- Similar to Fig.~6 but showing the errors in $Q_X$
for a special case, $\TsTg=1$, where an exact solution
for $Q_X$ exists (see \S4.1).
The mean and rms errors in \QX\ are $-5\times 10^{-4}$
and $5 \times 10^{-4}$, respectively.

\noindent
{\bf Fig.~9} --- (a) Solid curves: values of $R$ (top), \QJ\ (middle) and 
\QX\ (bottom) determined by numerical integration of the Langevin equations,
plotted vs.\ the axis ratio. Values of the dust-to-gas temperature ratio
and magnetic damping parameter are indicated.
Dashed curves: product of $\QX\QJ$ from
the numerical calculations (top), the exact value of \QJ\ for a spherical
grain with the same \TsTg\ and \deltm\ (middle) and the exact value of \QX\ for
a Maxwellian angular momentum distribution in the grain frame (bottom).
(b) Similar to (a) but for $\deltm=10$.

\noindent
{\bf Fig.~10} --- Similar to Figure~9 but showing the dependence on
the magnetic damping parameter with $a/b$ and $\TsTg$ fixed.

\noindent
{\bf Fig.~11} --- Similar to Figure~9 but showing the dependence on
the dust-to-gas temperature ratio with $a/b$ and $\deltm$ fixed.

\noindent
{\bf Fig.~12} --- Dependence of the correlation function
[see eq.~(\ref{eq-4_6})] 
on the magnetic damping parameter. Solid curve: numerical calculations with the
other parameters fixed at the indicated values.
Dashed curve: approximation to $\rho$ obtained by replacing the angular
momentum distribution with the angular momentum distribution for spheres with the same
\TsTg\ and \deltm\ (see text).

\noindent
{\bf Fig.~13} --- Comparison between the results of PS71 (filled circles)
and our numerical results (solid).
The dotted curves depict \QJ\ for spherical grains with the same
\deltm\ and \TsTg\ (middle panel) and \QX\ for a Maxwellian
angular momentum distribution in the body frame (bottom panel).

\noindent
{\bf Fig.~14} --- Similar to Fig.~13 but showing the dependence on \deltm.

%
%

%
%
\clearpage
\begin{figure}
\begin{picture}(450,550)
\includegraphics{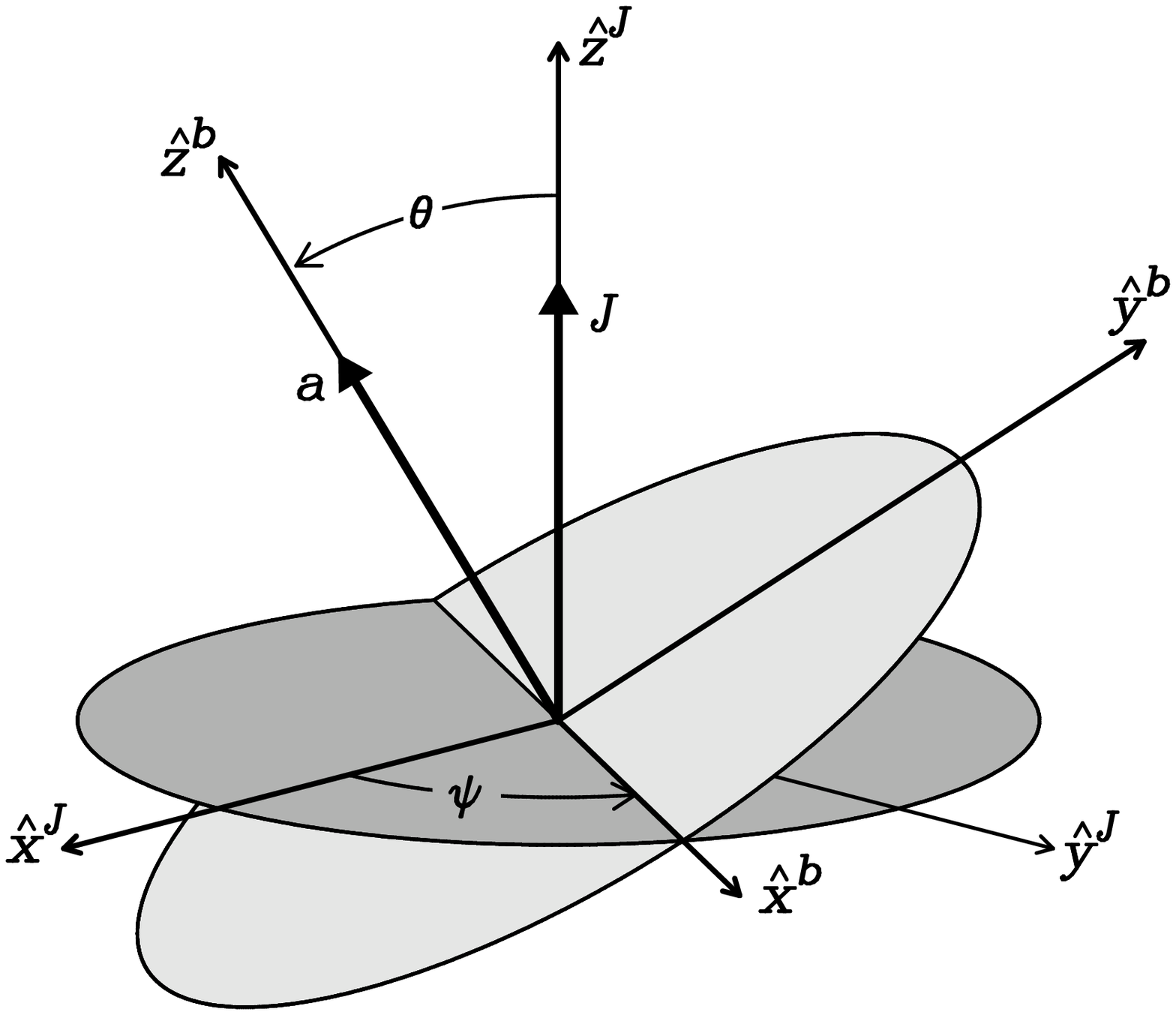}
\smallskip
\centerline{\large\bf Figure\ 1}
\end{picture}
\end{figure}

%
%
\clearpage
\begin{figure}
\begin{picture}(450,550)
\includegraphics{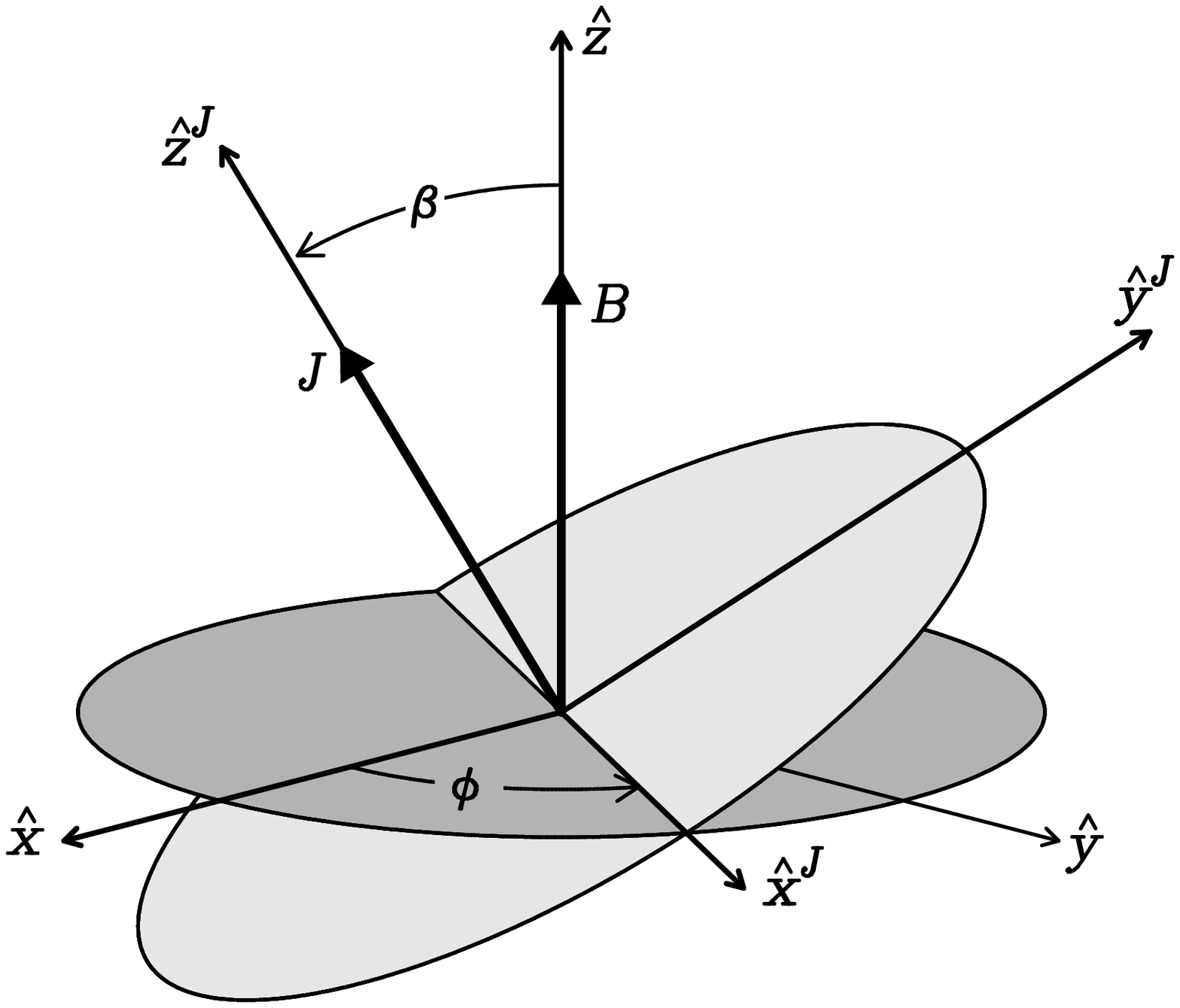}
\smallskip
\centerline{\large\bf Figure\ 2}
\end{picture}
\end{figure}

%
%
\clearpage
\begin{figure}
\begin{picture}(450,550)
\includegraphics{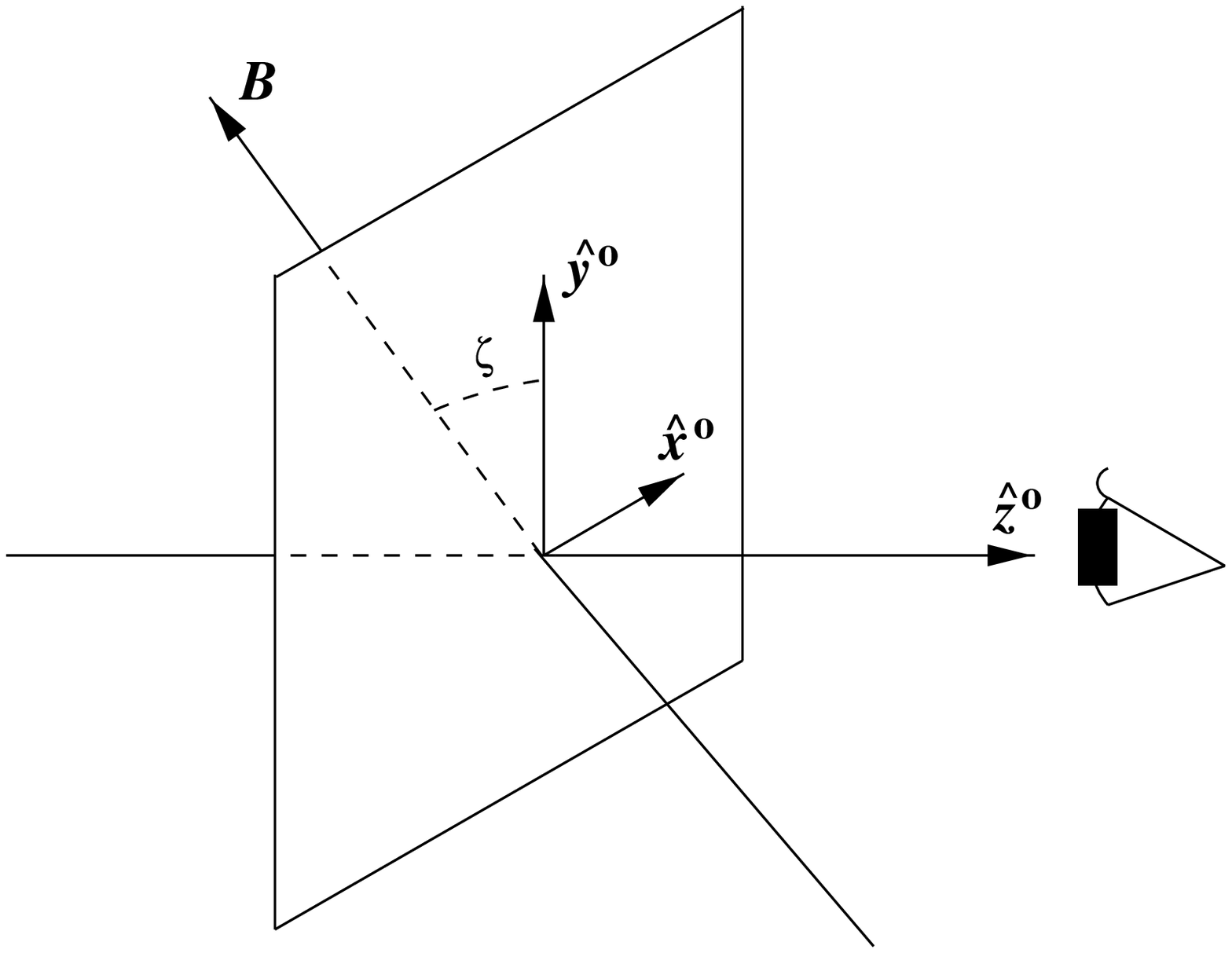}
\smallskip
\centerline{\large\bf Figure\ 3}
\end{picture}
\end{figure}

%
%
\clearpage
\begin{figure}
\begin{picture}(450,550)
\includegraphics{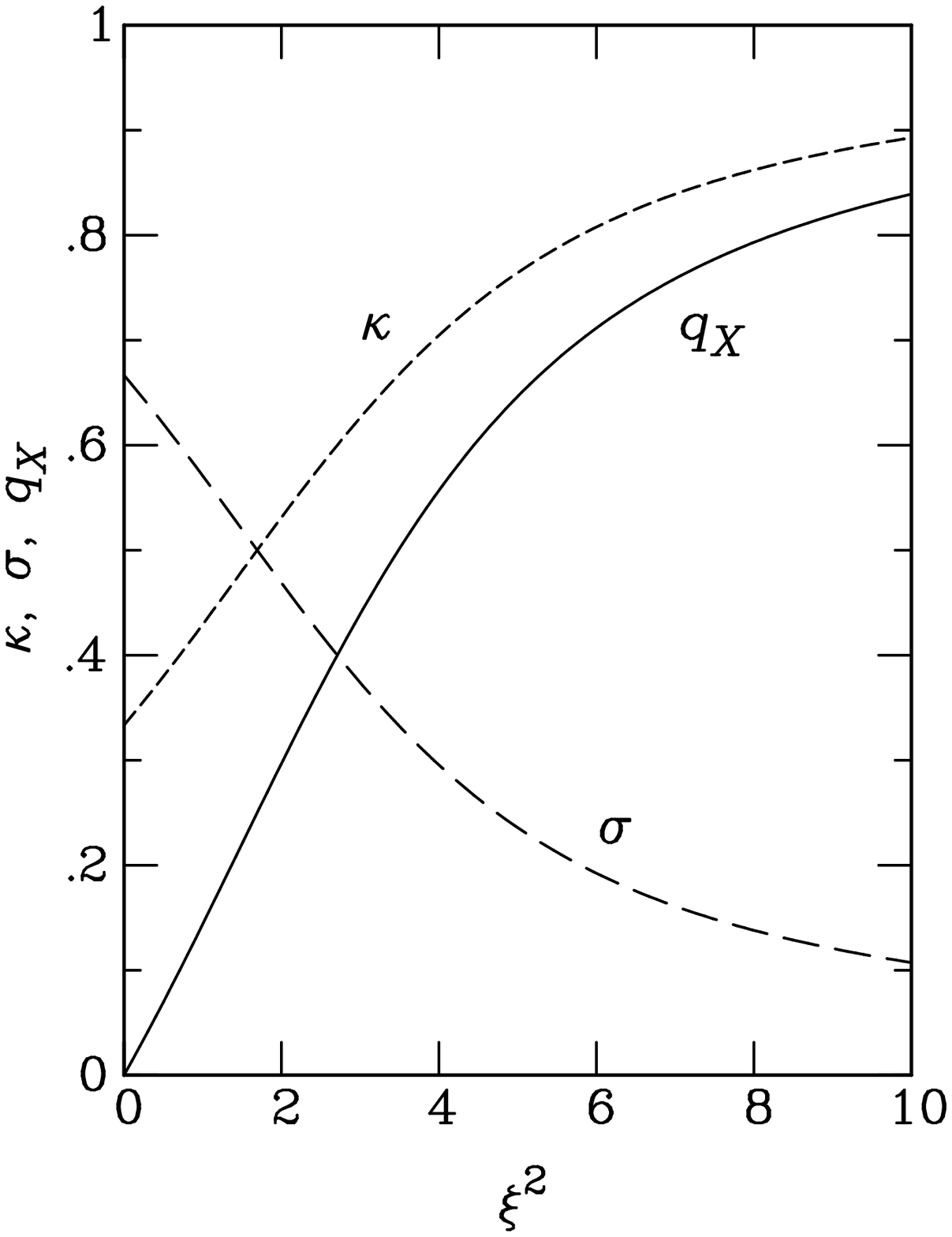}
\smallskip
\centerline{\large\bf Figure\ 4}
\end{picture}
\end{figure}

%
%
\clearpage
\begin{figure}
\begin{picture}(450,550)
\includegraphics{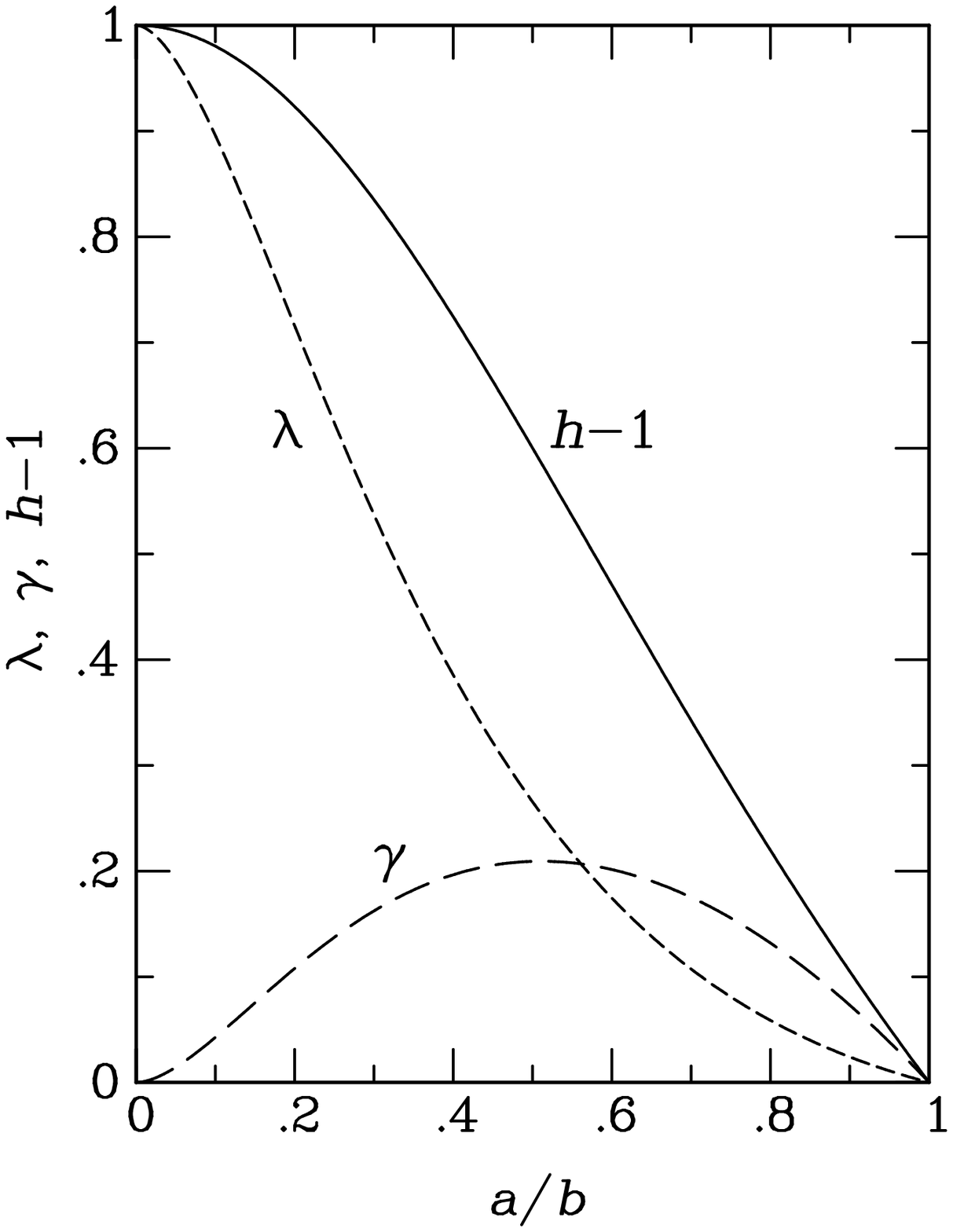}
\smallskip
\centerline{\large\bf Figure\ 5}
\end{picture}
\end{figure}

%
%
\clearpage
\begin{figure}
\begin{picture}(450,550)
\includegraphics{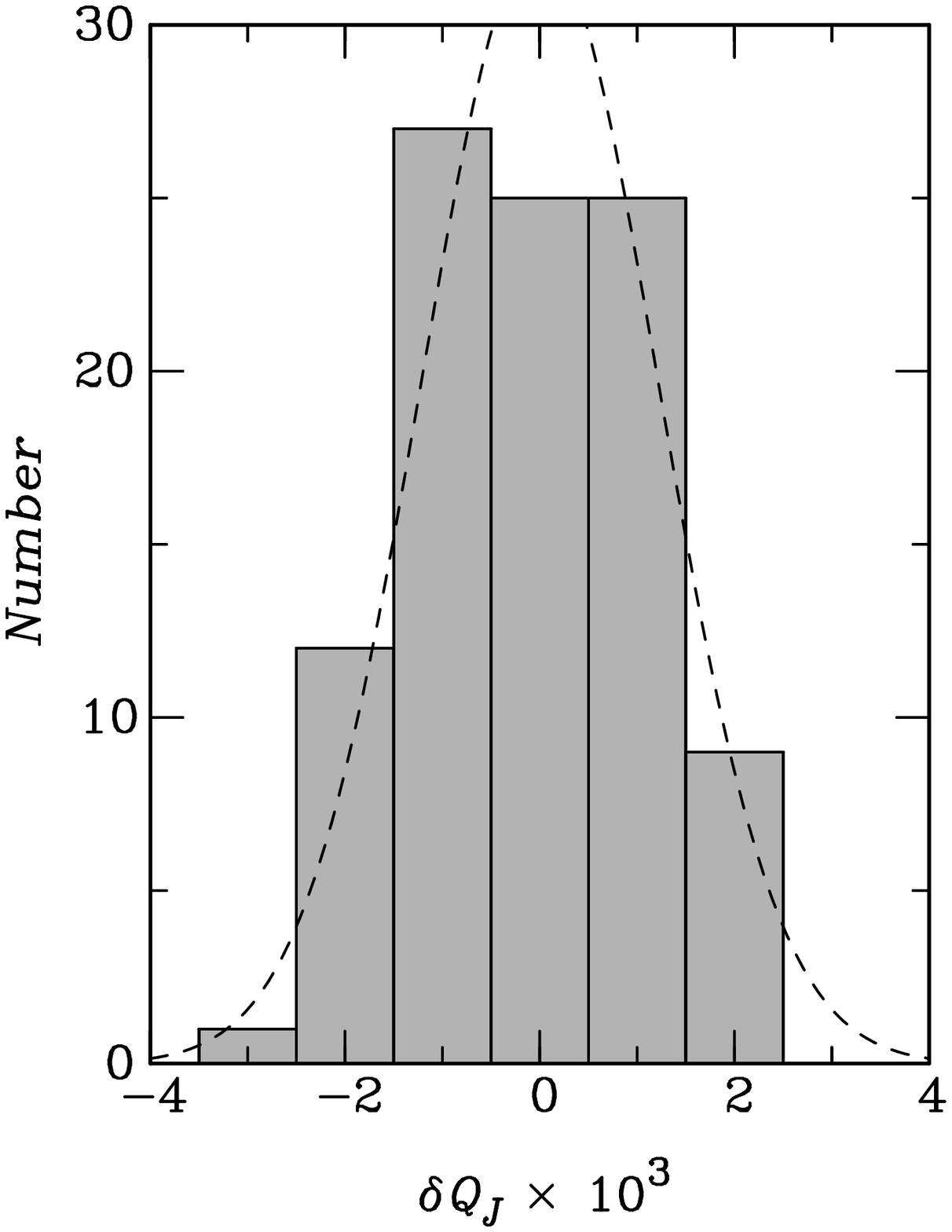}
\smallskip
\centerline{\large\bf Figure\ 6}
\end{picture}
\end{figure}

%
%
\clearpage
\begin{figure}
\begin{picture}(450,550)
\includegraphics{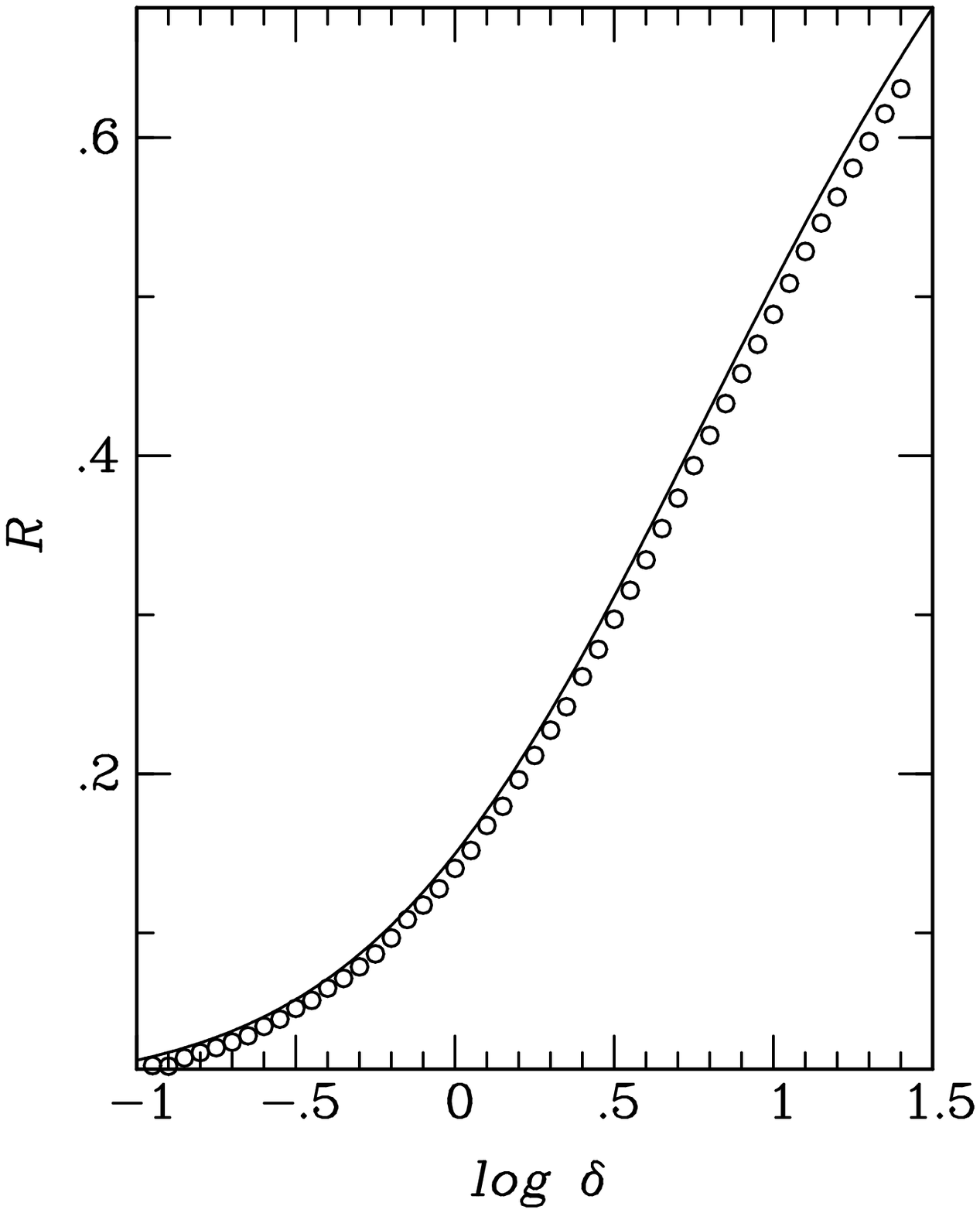}
\smallskip
\centerline{\large\bf Figure\ 7}
\end{picture}
\end{figure}

%
%
\clearpage
\begin{figure}
\begin{picture}(450,550)
\includegraphics{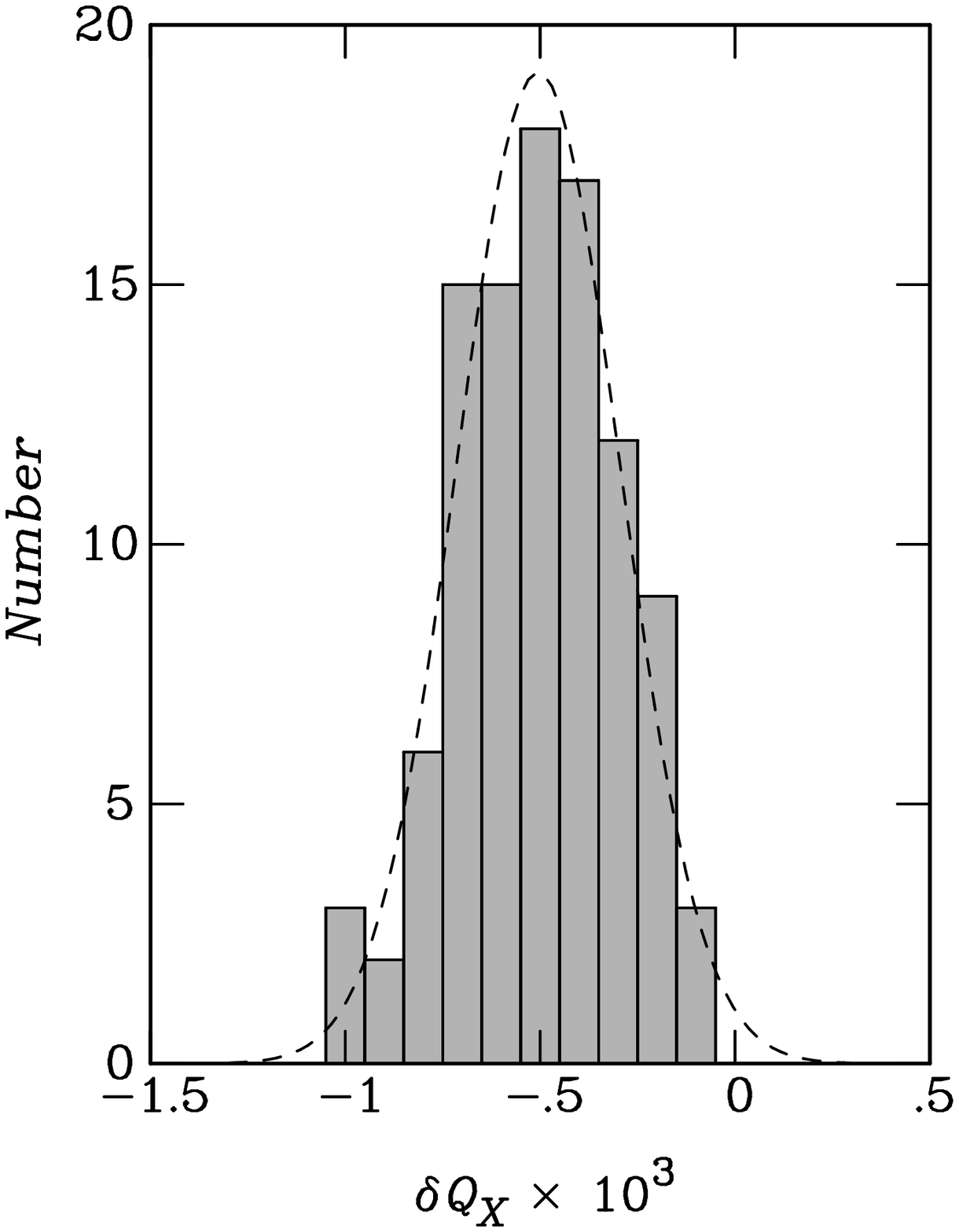}
\smallskip
\centerline{\large\bf Figure\ 8}
\end{picture}
\end{figure}

%
%
\clearpage
\begin{figure}
\begin{picture}(450,550)
\includegraphics{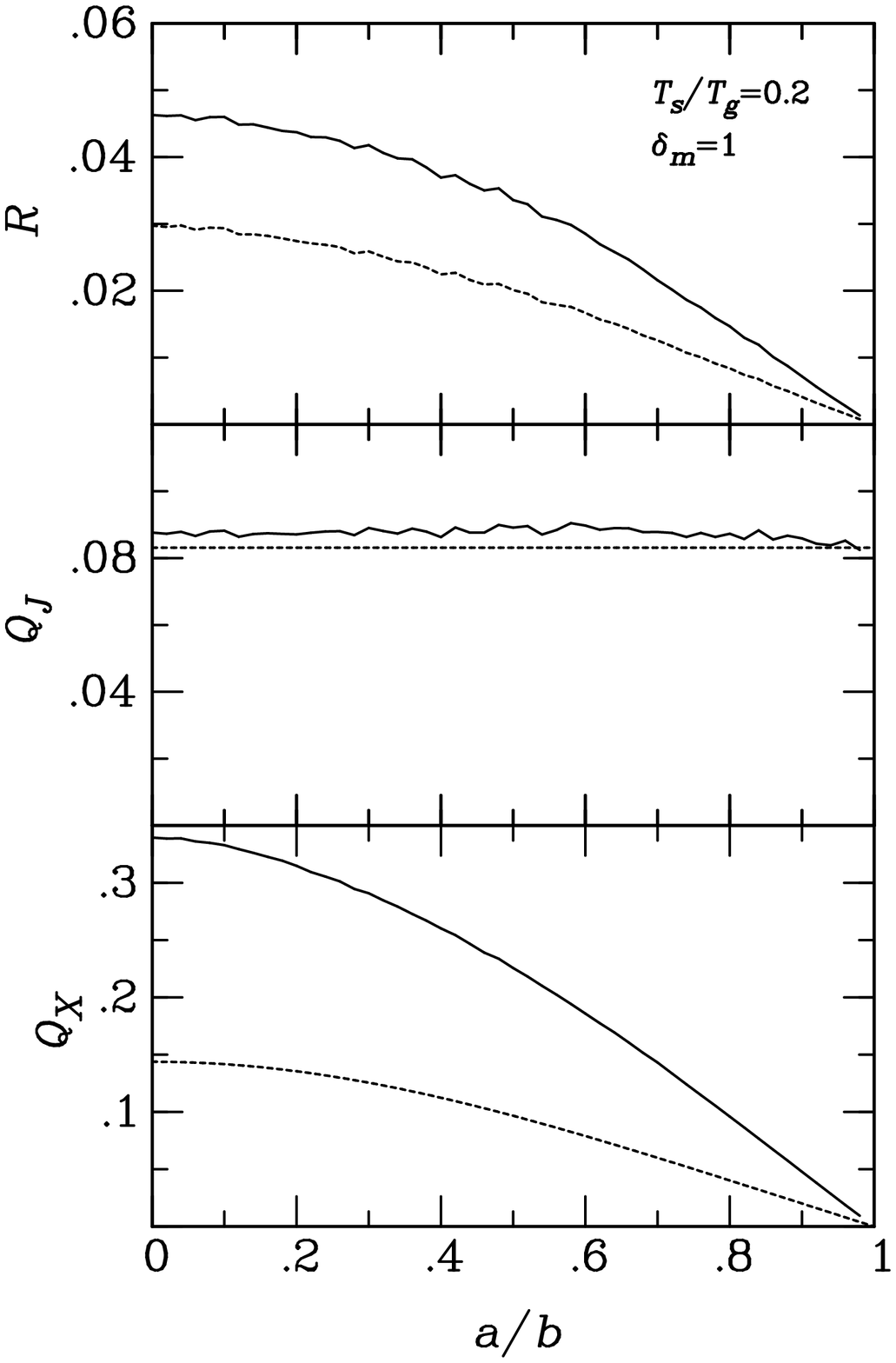}
\smallskip
\centerline{\large\bf Figure\ 9a}
\end{picture}
\end{figure}

%
%
\clearpage
\begin{figure}
\begin{picture}(450,550)
\includegraphics{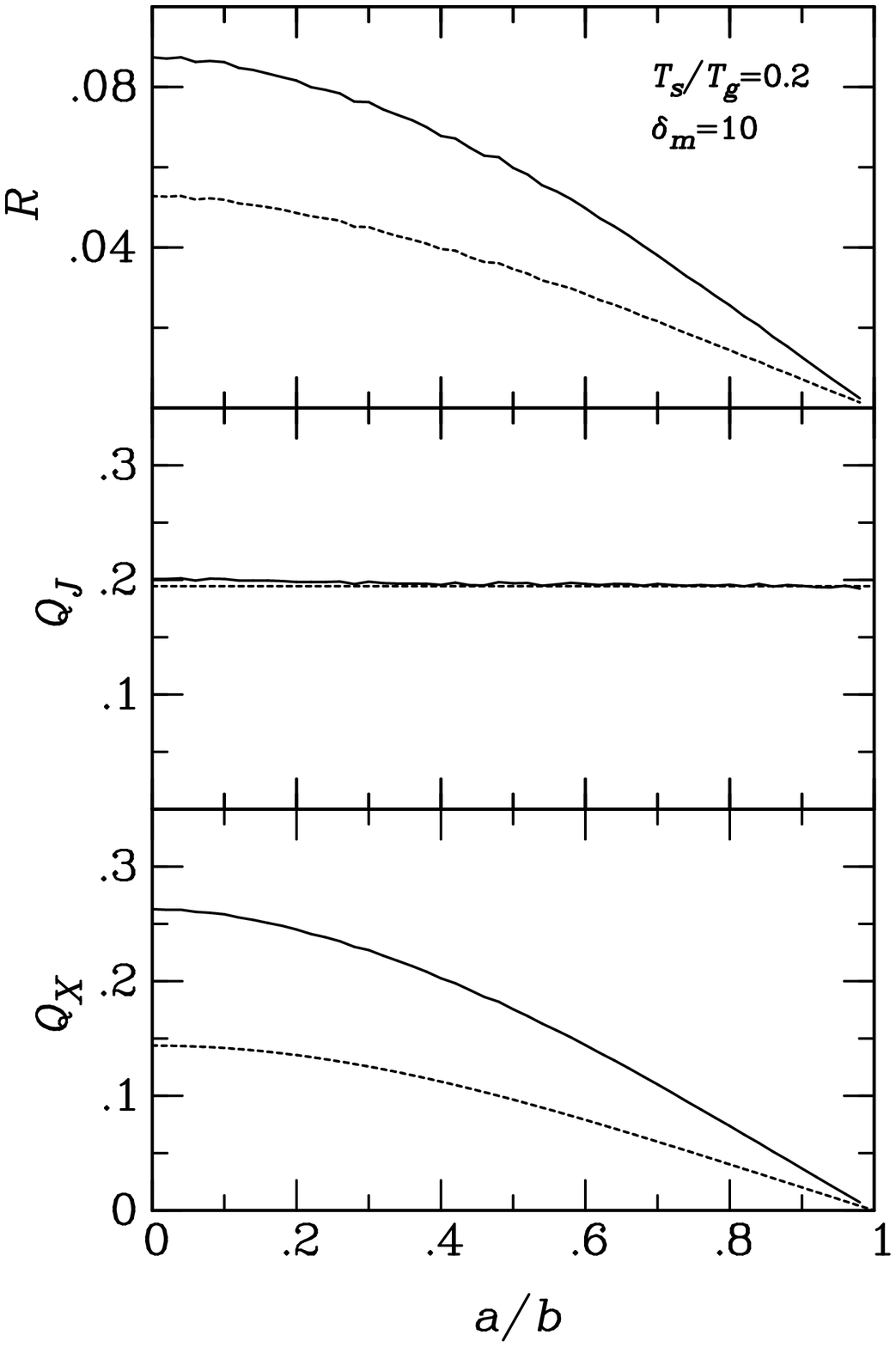}
\smallskip
\centerline{\large\bf Figure\ 9b}
\end{picture}
\end{figure}

%
%
\clearpage
\begin{figure}
\begin{picture}(450,550)
\includegraphics{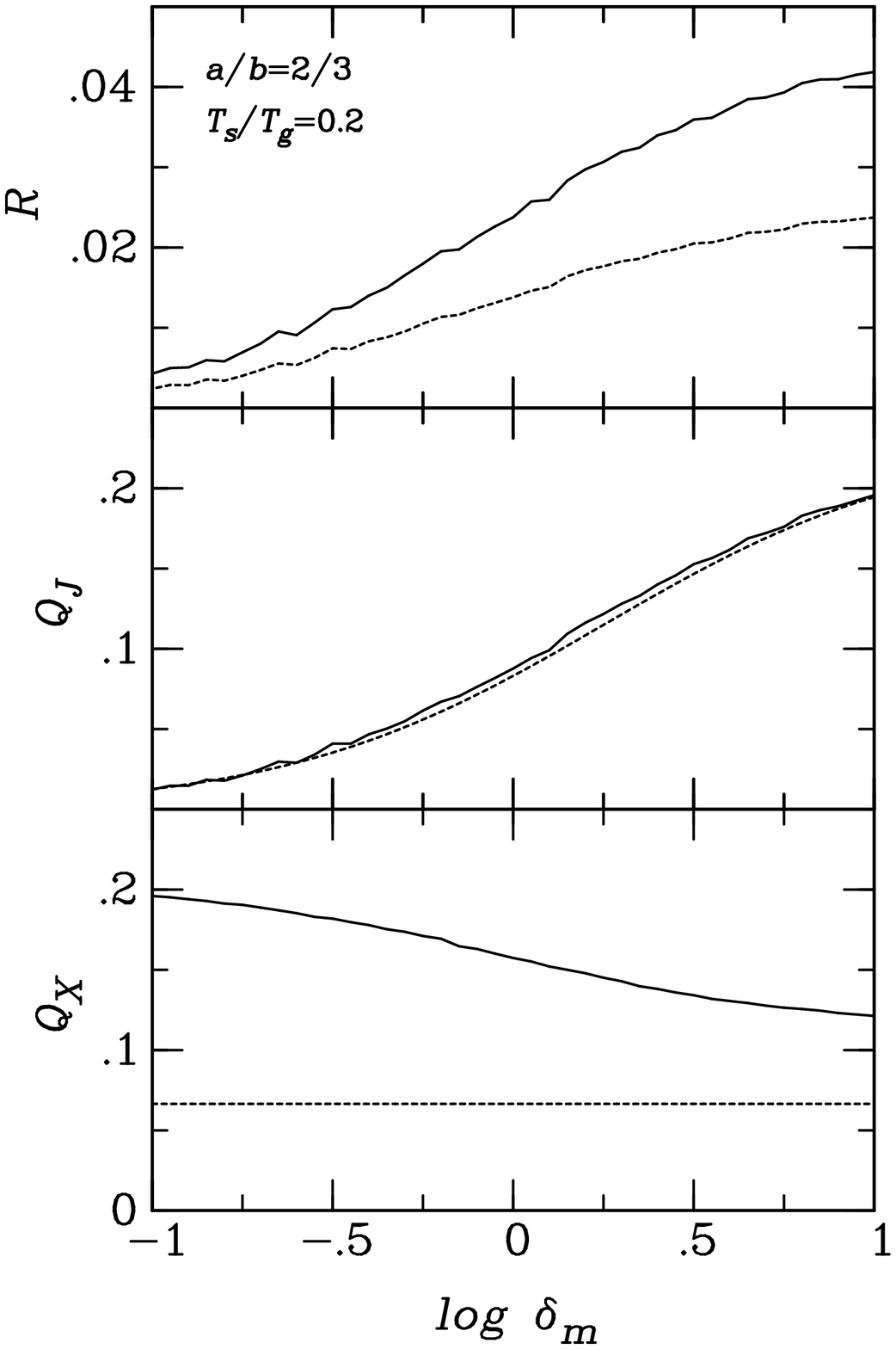}
\smallskip
\centerline{\large\bf Figure\ 10}
\end{picture}
\end{figure}

%
%
\clearpage
\begin{figure}
\begin{picture}(450,550)
\includegraphics{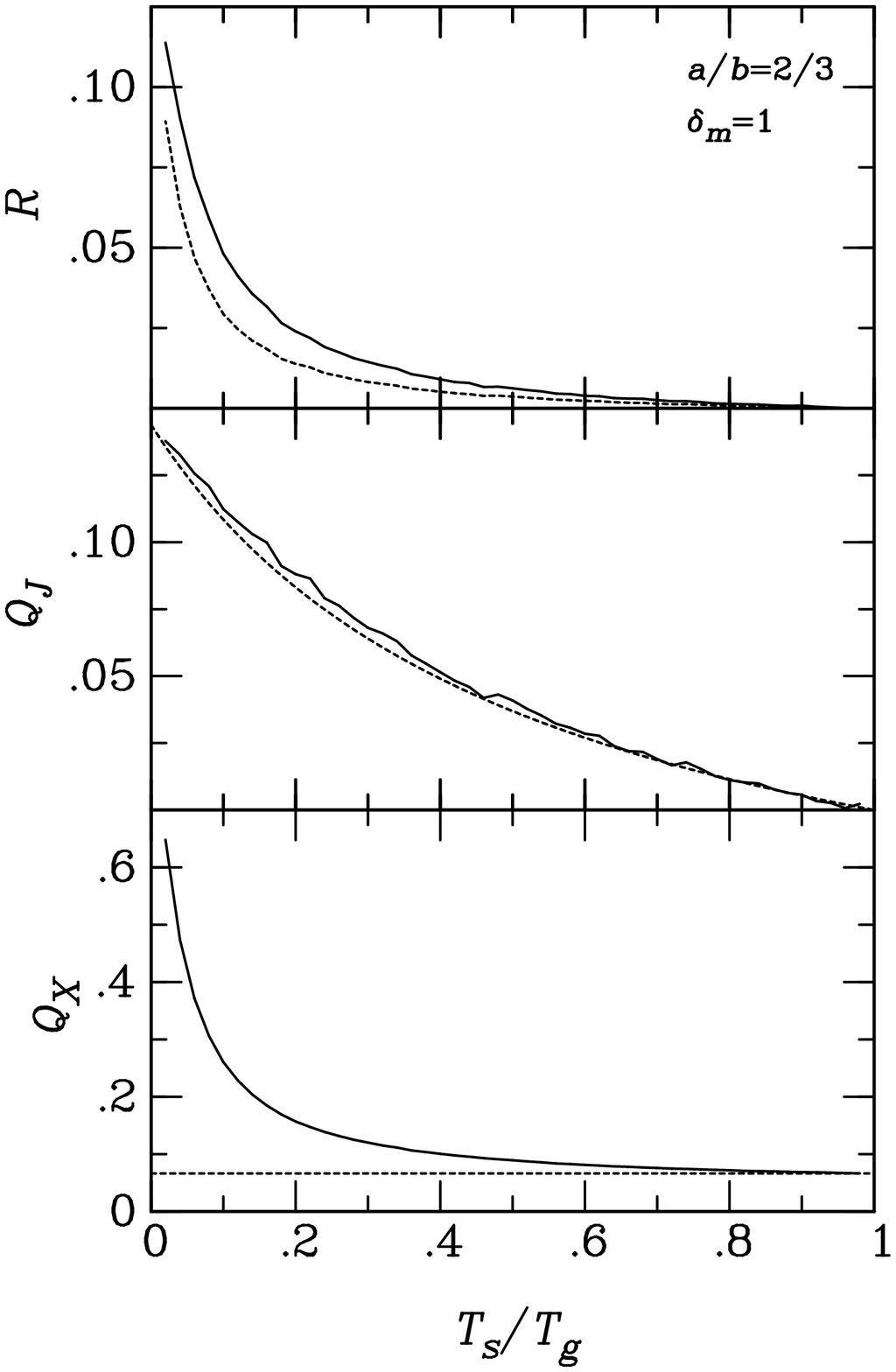}
\smallskip
\centerline{\large\bf Figure\ 11a}
\end{picture}
\end{figure}

%
%
\clearpage
\begin{figure}
\begin{picture}(450,550)
\includegraphics{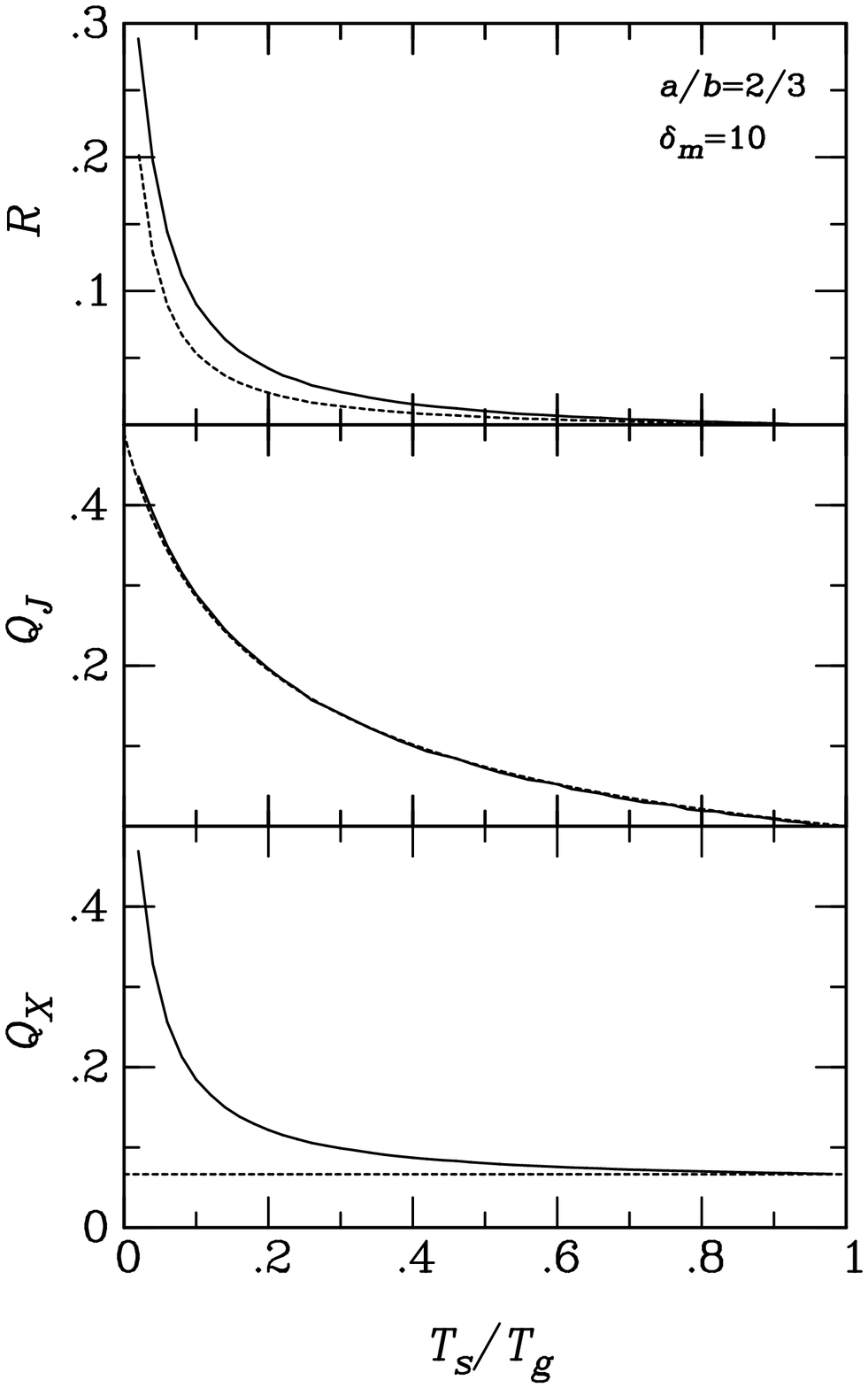}
\smallskip
\centerline{\large\bf Figure\ 11b}
\end{picture}
\end{figure}

%
%
\clearpage
\begin{figure}
\begin{picture}(450,550)
\includegraphics{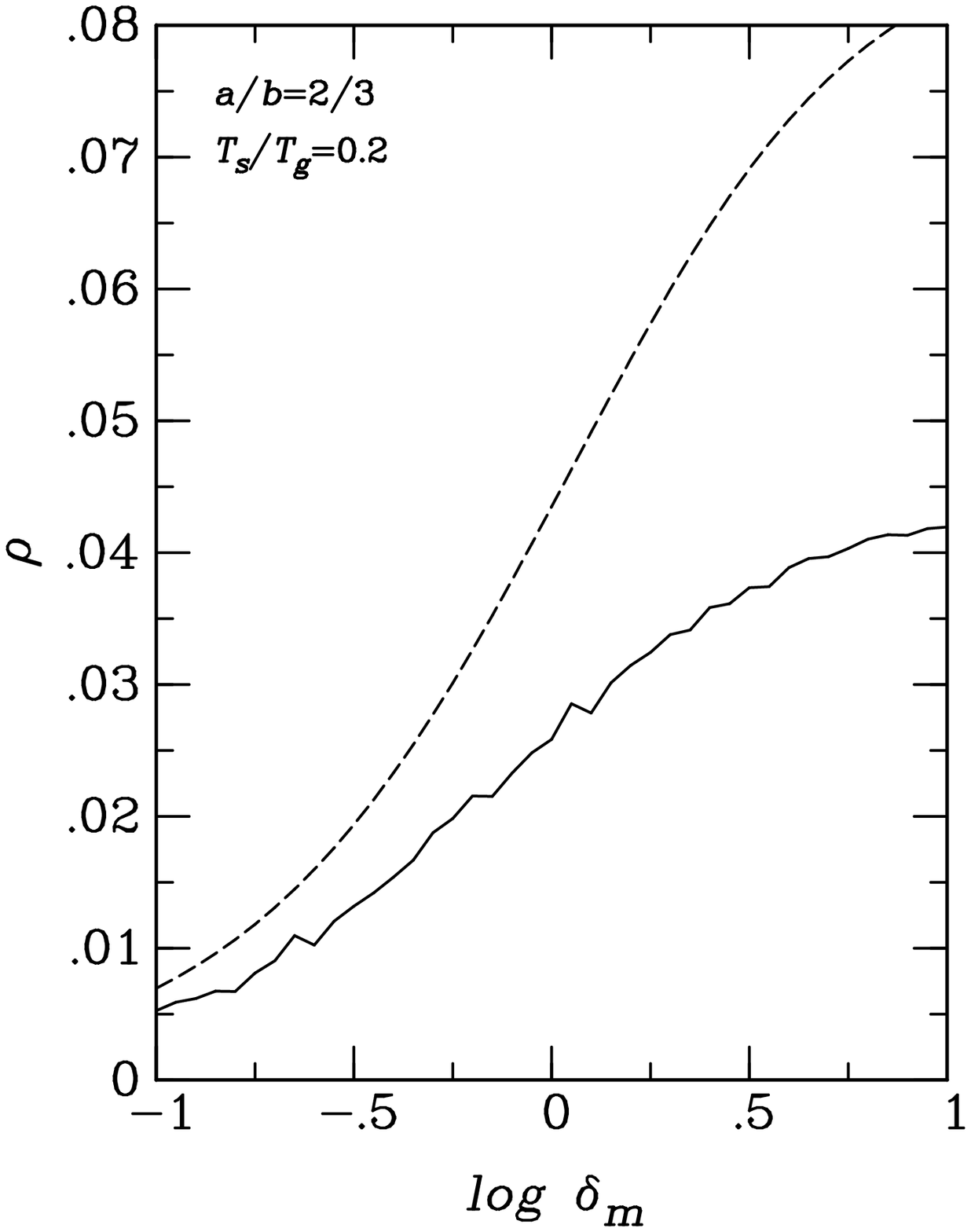}
\smallskip
\centerline{\large\bf Figure\ 12}
\end{picture}
\end{figure}

%
%
\clearpage
\begin{figure}
\begin{picture}(450,550)
\includegraphics{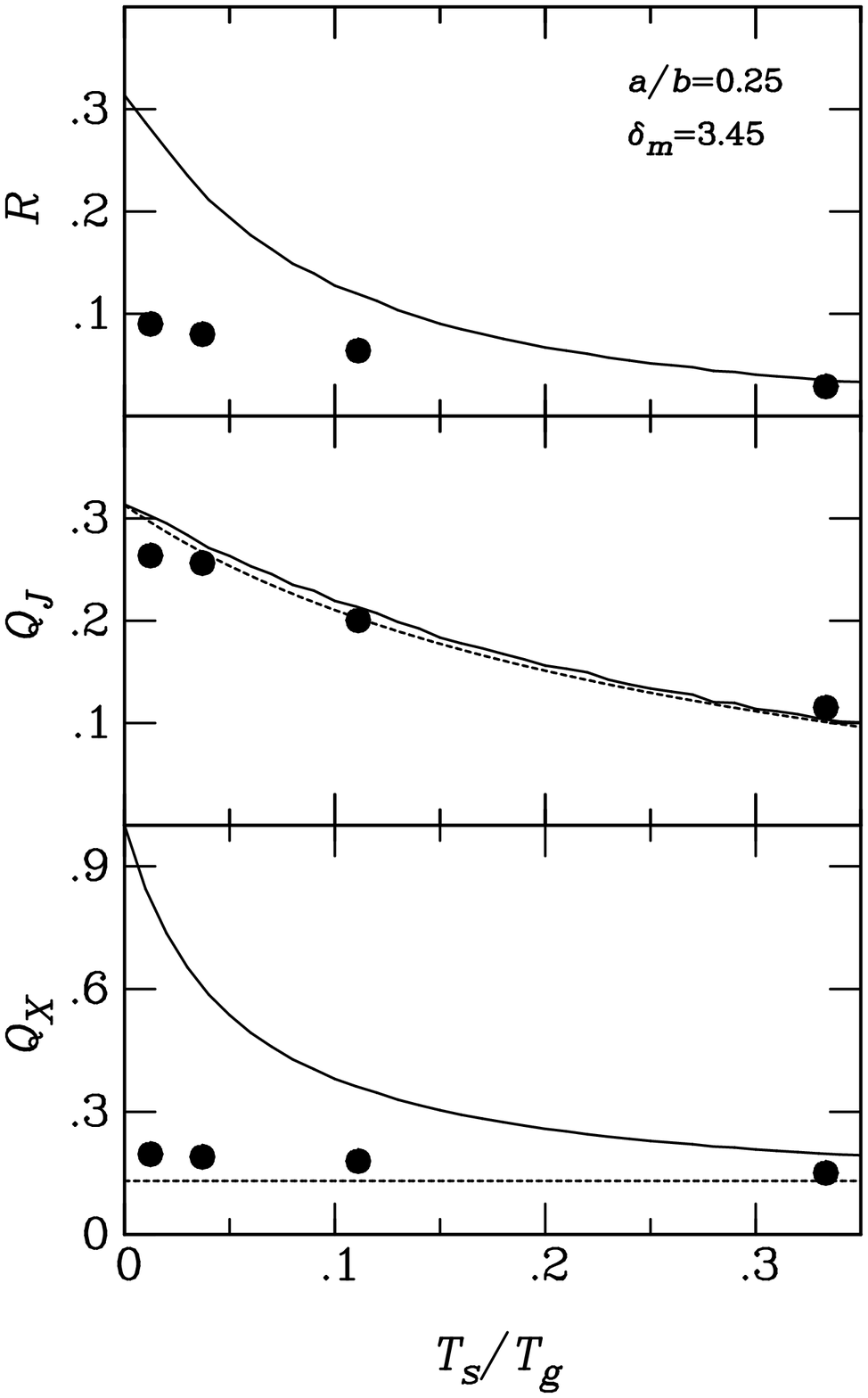}
\smallskip
\centerline{\large\bf Figure\ 13}
\end{picture}
\end{figure}

%
%
\clearpage
\begin{figure}
\begin{picture}(450,550)
\includegraphics{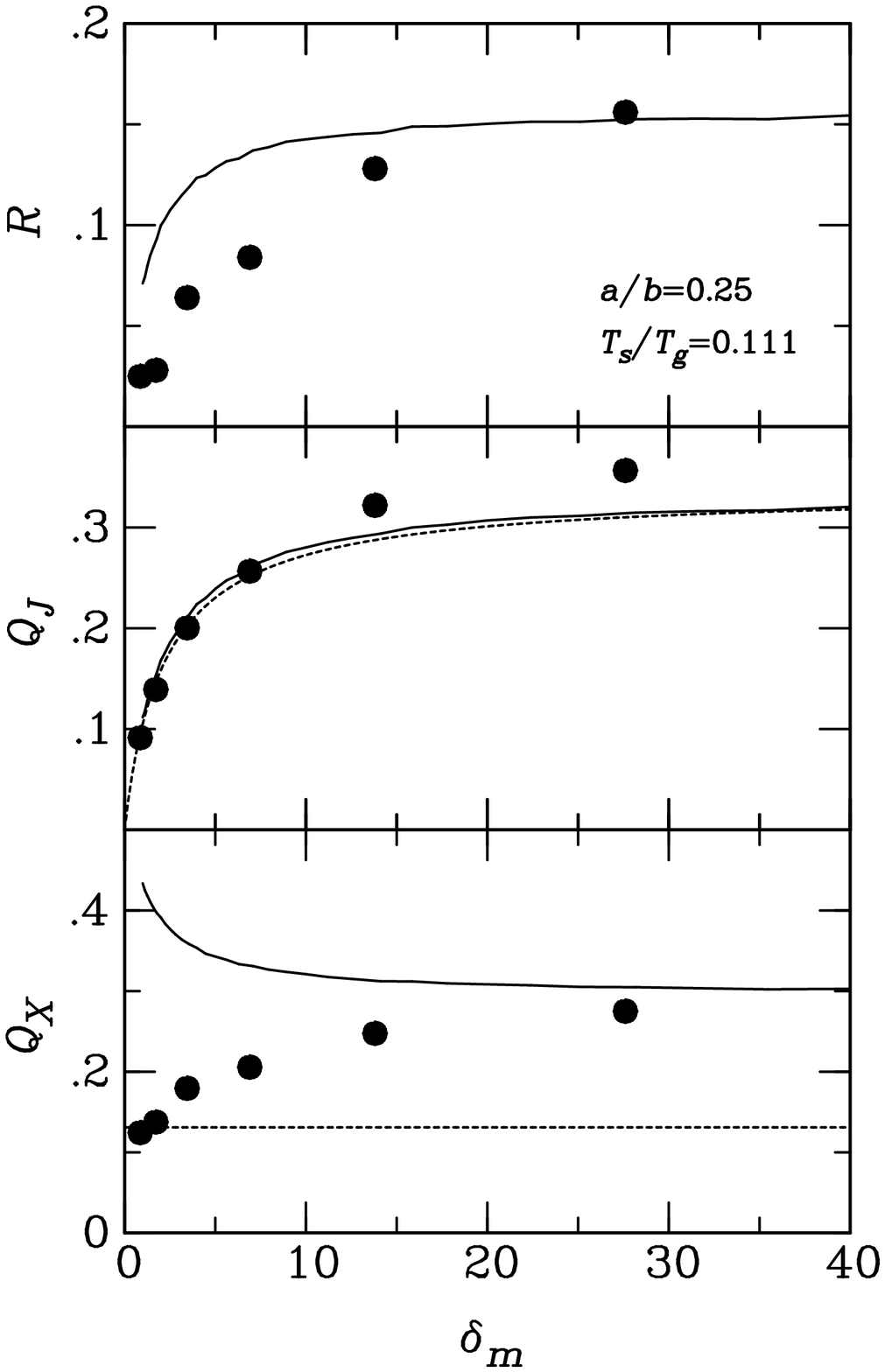}
\smallskip
\centerline{\large\bf Figure\ 14}
\end{picture}
\end{figure}

%
%

\newpage
\begin{thebibliography}{}
\bibitem{B94} Bradley J.P., 1994, Science, 265, 925
\bibitem{DG51} Davis L., Jr, Greenstein J.L., 1951, \apj, 114, 206 (DG51)
\bibitem{D96} Draine B.T., 1996, in Roberge W.G., Whittet D.C.B., eds,
   ASP Conf.\ Ser.\ Vol.\ 97, Polarimetry of the Interstellar Medium.
   Astron.\ Soc.\ Pac., San Francisco, p.\ 16
\bibitem{DL98} Draine B.T., Lazarian A., 1998, astro-ph 9807009 (DL98)
\bibitem{GW95} Goodman A.A., Whittet D.C.B., 1995, ApJ, 455, L181
\bibitem{H88} Hildebrand R.H., 1988, QJRAS, 29, 327
\bibitem{H96} \sameauthor\ 1996, in Roberge W.G., Whittet D.C.B., eds,
   ASP Conf.\ Ser.\ Vol.\ 97, Polarimetry of the Interstellar Medium.
   Astron.\ Soc.\ Pac., San Francisco, p.\ 254
\bibitem{JS67} Jones R.V., Spitzer, L., Jr, 1967, \apj, 147, 943 (JS67)
\bibitem{L94} Lazarian A., 1994, \mnras, 268, 713 (L94)
\bibitem{L95} Lazarian A., 1995, \apj, 453, 229 (L95)
\bibitem{L97} Lazarian A., 1997, \mnras, 288, 609 (L97)
\bibitem{L98} Lazarian A., 1998, \mnras, 293, 208
\bibitem{LE99} Lazarian A., Efroimsky, M. 1999, preprint
\bibitem{LR97} Lazarian A., Roberge W.G., 1997, \apj, 484, 230 (LR97)
\bibitem{LD85} Lee H.M., Draine B.T., 1985, \apj, 290, 211
\bibitem{M74} Martin P.G., 1974, \apj, 187, 461
\bibitem{M95} Martin P.G., 1995, \apj, 445, L63
\bibitem{M86} Mathis J.S., 1986, \apj, 308, 281
\bibitem{P69} Purcell E.M.,  1969, Physica, 41, 100
\bibitem{P79} \sameauthor\  1979, \apj, 231, 404 (P79)
\bibitem{PS71} Purcell E.M., Spitzer, L., Jr, 1971, \apj, 167, 31 (PS71)
\bibitem{R84} Risken H. 1984, The Fokker-Planck Equation, Springer-Verlag,
   1984
\bibitem{R97} Roberge W.G., 1997, \mnras, 291, 345 (R97)
\bibitem{RDGF93} Roberge W.G., DeGraff T.A., Flaherty J.E., 1993
           \apj, 418, 287 (RDGF93)
\bibitem{ST58} Spitzer, L., Jr, Tukey, J.W., 1951, ApJ, 114, 187
\bibitem{W97} Wolff M.J., Clayton G.C., Kim S.-H., Martin P.G., Anderson C.M.,
           1997, \apj, 478, 395
\end{thebibliography}
\end{document}